\documentclass[final]{agujournal2019}
\usepackage{url}
\usepackage{lineno}
\usepackage{amsmath,amssymb}
\usepackage{graphicx}
\usepackage{subcaption}
\usepackage{longtable}
\graphicspath{{figures/}}
\draftfalse
\journalname{Space Weather}
\begin{document}
\title{Forecasting the Ambient Solar Wind With a Physics-Informed Learned Inner Boundary: Part I—Differentiable Inverse Reconstruction From Solar Images}
\authors{Enrico Camporeale\affil{1,2}, Andong Hu\affil{2}, Prateek Mayank\affil{2}, Thomas E. Berger\affil{3}}
\affiliation{1}{Queen Mary University of London, UK}
\affiliation{2}{Space Weather Technology Research Education, (SWx-TREC), University of Colorado, Boulder, CO, USA}
\affiliation{3}{National Center for Atmospheric Research, High Altitude Observatory}
\correspondingauthor{Enrico Camporeale}{enrico.camporeale@qmul.ac.uk}
\begin{keypoints}
\item A differentiable heliospheric transport model enables inverse training of a solar-image-conditioned inner-boundary speed.
\item Skill comes from the speed-dependent travel time; a linear map matches the neural network, a four-coefficient model nearly so
\item More than half of the hourly speed variance is not determined by these inputs, so progress needs new information
\end{keypoints}

\begin{abstract}
The solar-wind speed at the inner boundary of heliospheric models is not observed; near-Earth monitors record the wind only after nonlinear propagation. We pose the reconstruction of an effective boundary speed at $21.5\,R_\odot$ as an inverse problem and train a map from solar observations to that boundary by differentiating through a reduced heliospheric transport solver from 21.5$\,R_\odot$ to measurement locations near the L1 Lagrangian point. Screening 42 candidate features from SDO/AIA 193~\AA{} images and SDO/HMI magnetograms identifies coronal-hole area, its proximity to the ecliptic and its darkness as the most informative; the reference model adds their recent history, the expansion factor of GONG-based potential-field models and the speed observed one rotation earlier. In five-fold chronological cross-validation over solar cycle 24, the reconstruction outperforms 27-day recurrence and a linear regression on the screened features at a fixed lead (ROC area for wind above 600~km~s$^{-1}$: 0.879 versus 0.802 and 0.850) with a predicted-to-observed standard-deviation ratio of 0.81, at the cost of a $+26$~km~s$^{-1}$ bias from its tail-weighted loss. A factorial comparison shows that the skill comes from training through a speed-dependent propagation: a fixed lag degrades the score by 8.6~km~s$^{-1}$, stream interaction adds little, and a linear map, or a model with one coefficient per physical quantity, matches the neural network. About 57\% of the variance of the hourly speed is not determined by these inputs, and the remaining error is common to all model classes: better reconstructions need more informative inputs rather than more flexible models. Forecasting and probabilistic prediction are addressed in Parts II and III.
\end{abstract}

\section*{Plain Language Summary}
Models of the solar wind need to know how fast it leaves the Sun, but spacecraft measure it only after it has traveled over 200 solar radii to near Earth locations. We connect possible speeds near the Sun to the speeds measured near Earth with a simple physical model, and train a machine learning (ML) model to infer the near-Sun speeds from solar images. Dark regions called coronal holes, the sources of fast wind, provide the most useful data, together with the speed measured one solar rotation earlier. What matters most is that the model accounts for the fact that the time the wind takes to reach Earth depends on its speed; a simple formula with four ingredients works as well as a neural network. Even so, very similar solar conditions are sometimes followed by fast wind and sometimes by slow wind, so no more than half of the hour-to-hour variation can be predicted from these observations. Better predictions will need new kinds of observations rather than more complex models. Forecasting from a fixed time, and attaching probabilities to the predictions, are the subjects of follow-up studies.

\section{Introduction}
\label{sec:intro}

The solar wind is a continuous, supersonic outflow of magnetized plasma from the solar corona \cite{Parker1958}. Its acceleration is largely complete within a few tens of solar radii, beyond which it expands nearly radially and carries the coronal magnetic field outward, wound into a spiral by solar rotation \cite{Parker1958,WeberDavis1967}. Near Earth its speed ranges from roughly 300 to more than 700~km\,s$^{-1}$. Fast wind flows from the open magnetic field of coronal holes, whereas slow wind is associated with coronal-hole boundaries and the streamer belt, although its origin remains debated. Because the Sun rotates, sources of different speed face a given direction in turn, so fast wind runs into slower wind emitted ahead of it. The resulting compressions form corotating interaction regions (CIRs), which recur with the roughly 27-day solar rotation and shape the plasma and magnetic field encountered near Earth \cite{Richardson2018,Verscharen2019, mayank2026finding}. Solar-wind speed is therefore central to space-weather prediction in two ways: it sets the ambient flow through which coronal mass ejections (CMEs) propagate, and it drives recurrent high-speed-stream effects at Earth.

Errors in the ambient wind propagate into predicted CME arrival times \cite{Parsons2011}: in one sensitivity study, lowering the background speed by 25\% delayed the modelled arrival by 12~h and raising it by 25\% advanced it by 7~h \cite{Kay2020}. Independently of CMEs, high-speed streams and the CIRs at their leading edges sustain geomagnetic activity over several days, and account for most of the geomagnetic indices outside solar maximum \cite{Richardson2000,Richardson2006}. Speed is not the only driver, since storm strength also depends on the orientation of the magnetic field \cite{Newell2016}, but better speed estimates support both CME-arrival prediction and the assessment of these recurrent drivers.

Because the solar wind is not monitored in situ close to its source, heliospheric models start from an inner boundary, typically at $21.5\,R_\odot$ (0.1~AU), where the speed must be prescribed from remote observations of the Sun. The standard input is the photospheric magnetic field. The Global Oscillation Network Group (GONG), a ground-based network of six observatories, measures it continuously, and its full-disk magnetograms are assembled into synoptic maps of the whole solar surface \cite{Harvey1996}. The coronal field is extrapolated from such maps with the potential-field source-surface (PFSS) model, which assumes a current-free corona out to a spherical source surface, usually at $2.5\,R_\odot$, beyond which the field is radial \cite{Altschuler1969,Schatten1969}. Field lines traced in this model identify the open-field regions, the coronal holes, and how much each open flux tube expands. Empirical relations then convert this geometry into a speed: the Wang--Sheeley--Arge (WSA) model assigns faster wind to flux tubes that expand less and are rooted farther from a coronal-hole boundary \cite{WangSheeley1990,WangSheeley1992,Arge2000,Arge2003}. The resulting boundary map is finally propagated outward either by a full magnetohydrodynamic (MHD) model, such as ENLIL or EUHFORIA, or by a reduced model such as the Heliospheric Upwind eXtrapolation (HUX) model and its time-dependent extension HUXt, which keep only the radial advection of the speed and trade physical completeness for computational efficiency \cite{Riley2011,Owens2020}. The chain of magnetogram, PFSS extrapolation, WSA relation and MHD propagation is used operationally as WSA--ENLIL \cite{Odstrcil2003} at NOAA's Space Weather Prediction Center, where it entered operations in 2011 \cite{Parsons2011}, and at the UK Met Office \cite{Gonzi2021}, and as EUHFORIA \cite{Pomoell2018} within ESA's Virtual Space Weather Modelling Centre \cite{Poedts2020VSWMC}.

Solar-wind data assimilation uses in-situ measurements to update the heliospheric state or the inner-boundary map, an approach that has progressed from early ensemble filtering with ENLIL to the variational BRaVDA scheme and its operational extensions \cite{Lang2017,Lang2019,Lang2021,Turner2023,Turner2025}. Schemes such as BRaVDA update the inner boundary of a HUX-type model with the in-situ speeds observed over the preceding solar rotation, so the information they add about wind arriving at Earth in the next few days is about a rotation old; fresher in-situ information requires a monitor away from the Sun--Earth line, such as at L5. Remote sensing offers a complementary route: interplanetary scintillation tomography reconstructs large-scale heliospheric structure from remote observations \cite{Jackson1998}.

Machine learning offers a further route, learning a direct relation between solar observations and the wind at L1 from data. \citeA{Upendran2020} trained a deep network on sequences of AIA images to predict the daily speed at L1 several days ahead, and found that it attended to coronal holes when predicting fast wind and to active regions for slow wind. Later work matched each speed to its source image by ballistic back-tracing \cite{Raju2021} and replaced convolutional by attention-based architectures, which performed best in the declining phase of the solar cycle \cite{Brown2022}. Feature-based models reach comparable or better accuracy with far fewer parameters: a polynomial regression on gridded coronal-hole areas and the speed observed one rotation earlier outperformed deep networks, and showed that training on the mean squared error systematically underestimates the peak speeds of high-speed streams \cite{Collin2025}; a probabilistic extension predicts the full distribution of the speed \cite{Collin2026}. WSA+ combines learning with an empirical physical relation, using automatic differentiation to optimize the WSA formula against downstream measurements \cite{Mayank2025}. Most of these methods predict the speed at L1 from images taken a fixed time earlier, or matched to it by ballistic back-tracing. Propagation is not modelled explicitly (although one could argue that it might be implicitly learned by the neural network) and the product is a time series at one point rather than a boundary condition that a heliospheric model can carry to other locations or use as the ambient wind for CMEs.

In this paper, we formulate reconstruction of the solar-wind speed at the inner boundary as an inverse problem: infer a time-dependent boundary field $u(\phi,t)$ whose forward propagation is consistent with the solar-wind measurements at L1. The inference is ill-posed. A single time series at L1 reflects plasma launched from multiple longitudes and at different times, while nonlinear stream interactions mix contributions from neighboring sources; consequently, distinct boundary fields can produce similar measurements. We therefore regularize the reconstruction and constrain it with independent information about the solar source, while requiring consistency with a physically motivated forward model.

Solar images are the natural source of that information, because coronal holes, the sources of fast wind, are visible several days before their wind reaches Earth. The Empirical Solar Wind Forecast (ESWF) exploits this by relating the dark-pixel fraction in a central-meridian strip of extreme-ultraviolet images to the solar-wind speed observed about four days later \cite{Rotter2012,Rotter2015,Reiss2016}. However, the relation between image structure and downstream speed is noisy, and the observed structures are distorted by projection toward the limb, by instrument degradation, and by the evolution of the corona over the solar cycle. Whether images can constrain an inversion therefore depends on which image-derived quantities carry information about the wind, and on how stable that information is across viewing geometries and solar-cycle phases, all of which remain open research questions. We use this information by representing the boundary speed at every Carrington longitude and time as the output of one function of the observations of that longitude, using a neural network shared by all longitudes and times as a function approximators. Its parameters are fitted so that the boundary, propagated to L1, reproduces the measured speed; the boundary speed itself is of course never observed. Because fitting the mean squared error would compress the predicted amplitude of relatively rare high-speed streams, the fit uses a score that gives more weight to errors at faster observed speeds (Section~\ref{sec:metric}).

This paper is the first of a three-part series. Part~I formulates the reconstruction of the boundary as an inverse problem and asks what the solar observations can and cannot determine; its assessments are reconstructions, which use the information available when the relevant plasma leaves the inner boundary, and must not be confused with an operational forecast issued at a fixed time. Part~II turns the reconstruction into a deterministic forecast without future solar images, tests it on solar-cycle-25 data that were not used at any stage of development, compares it with established solar-wind forecasts, and presents a concept of operations for real-time use. Part~III addresses probabilistic prediction and the calibration of its uncertainty.

Part I addresses five questions. First, can a map from solar observations to the inner-boundary speed be learned from downstream measurements alone, by differentiating through the propagation from the boundary to ACE, and how does it compare with simple empirical methods and with the established physics-based chain, WSA propagated with HUXt, evaluated on the same hours? Second, which observations carry the information? We screen 42 candidate features from extreme-ultraviolet images (SDO/AIA, 193~\AA{}) and photospheric magnetograms (SDO/HMI), and use the flux-tube expansion factor of PFSS extrapolations of GONG maps, the magnetic ingredient of WSA; other sources can be added in the same framework. Third, which element of the end-to-end formulation produces the skill? A factorial comparison replaces, one at a time, the training loss, the propagation (by a fixed lag or by ballistic parcels) and the network (by simpler functions). Fourth, how far do these inputs determine the speed at all? We search for situations with similar inputs and estimate, by a bias--variance--noise decomposition, the part of the error that no model using these inputs can remove. 
Fifth, can the learned boundary be expressed in a few physical terms, along the lines of interpretable machine learning?

The answers, anticipated here, are consistent with one another. The boundary map can be learned through the propagation. It outperforms 27-day recurrence and the WSA--HUXt chain, and, compared with a conventional linear regression of image features at a fixed lag, it discriminates fast wind modestly better and reproduces much more of the observed variability, although its training score leaves it with a positive bias and a higher raw RMSE. This advantage does not come from the neural network: a linear function of the same inputs, trained in the same way through the propagation, does as well. It comes mainly from training through a propagation in which the travel time depends on the speed; replacing the transport solver by ballistic propagation, in which parcels travel at their own speed without interacting, changes the skill little for the smooth boundaries learned here, and a model with one coefficient for each of four physical quantities comes close. The inputs themselves are ambiguous: for about one fast stream in six, the most similar situations in other years were followed by slow wind, and nearly three quarters of the model's squared error is estimated to be noise given its inputs. A substantially more accurate reconstruction will therefore need additional or different information rather than a more flexible map.

\section{Differentiable propagation and inverse formulation}
\label{sec:forward}
\subsection{Reduced Radial Transport}
\label{sec:transport}

We model propagation in the ecliptic plane beyond the principal acceleration region, using cylindrical coordinates $(r,\varphi,z)$ in an inertial frame, with $\varphi$ the inertial azimuth. The starting point is the ideal-MHD momentum equation,
\begin{equation}
\rho\left(\frac{\partial\mathbf{v}}{\partial t}+(\mathbf{v}\cdot\nabla)\mathbf{v}\right)
=-\nabla p+\frac{1}{\mu_0}(\nabla\times\mathbf{B})\times\mathbf{B}+\rho\mathbf{g},
\label{eq:mhd_momentum}
\end{equation}
where $\rho$ is mass density, $\mathbf{v}$ is inertial-frame velocity, $p$ is thermal pressure, $\mathbf{B}$ is the magnetic field, and $\mathbf{g}$ is gravitational acceleration. Restricting to a two-dimensional ecliptic flow, the radial component is
\begin{equation}
\frac{\partial v_r}{\partial t}+v_r\frac{\partial v_r}{\partial r}
+\frac{v_{\varphi}}{r}\frac{\partial v_r}{\partial\varphi}
-\frac{v_{\varphi}^2}{r}
=-\frac{1}{\rho}\frac{\partial p}{\partial r}
+\frac{1}{\mu_0\rho}\left[(\nabla\times\mathbf{B})\times\mathbf{B}\right]_r+g_r,
\label{eq:radial_momentum_full}
\end{equation}
where $-v_{\varphi}^2/r$ is the centripetal term that arises from the azimuthal variation of the radial unit vector.

Beyond the acceleration region, we neglect radial pressure gradients, gravity, and Lorentz forces. We also take the inertial azimuthal speed to be small compared with the radial speed, $v_{\varphi}\simeq0$, and set $v_z=0$, which removes the azimuthal-advection and centripetal terms. The radial speed then obeys the inviscid Burgers equation along each inertial azimuth,
\begin{equation}
\frac{\partial v_r}{\partial t}+v_r\frac{\partial v_r}{\partial r}=0.
\label{eq:huxt_radial}
\end{equation}
This is the radial equation solved by HUXt, in which solar rotation is represented geometrically through the longitude mapping of the boundary \cite{Owens2020}. Each azimuth evolves independently, which makes the inertial form computationally efficient and embarrassingly parallel.

\subsection{Co-Rotating Formulation}
\label{sec:corotating}

We solve the same transport in a co-rotating frame, which places the full boundary field on a fixed grid. Let $\phi$ denote the azimuth in a frame rotating with the Sun at the constant angular speed $\Omega_\odot$,
\begin{equation}
\phi=\varphi-\Omega_\odot t,
\label{eq:rotating_frame}
\end{equation}
with $r$, $z$ and $t$ unchanged. With $\Omega_\odot$ equal to the sidereal rotation rate that defines the Carrington coordinate system, $2.8653\times10^{-6}$~rad~s$^{-1}$ (a period of 25.38~d), $\phi$ is Carrington longitude, up to a constant. The radial velocity is the same in both frames, but its partial time derivative is not: the inertial derivative holds $\varphi$ fixed, whereas the rotating one holds $\phi$ fixed. Since $\varphi=\phi+\Omega_\odot t$, the chain rule gives
\begin{equation}
\left.\frac{\partial v_r}{\partial t}\right|_{\varphi}
=\left.\frac{\partial v_r}{\partial t}\right|_{\phi}
-\Omega_\odot\frac{\partial v_r}{\partial\phi},
\label{eq:time_derivative_transform}
\end{equation}
and Equation~\ref{eq:huxt_radial} becomes, writing $\Omega$ for $\Omega_\odot$ and leaving the fixed coordinate implicit,
\begin{equation}
\frac{\partial v_r}{\partial t}+v_r\frac{\partial v_r}{\partial r}-\Omega\frac{\partial v_r}{\partial\phi}=0.
\label{eq:forward}
\end{equation}
The longitude derivative expresses the apparent motion of radially moving parcels, which do not rotate with the Sun and therefore drift toward decreasing Carrington longitude at the rate $\Omega$. The numerical grid is uniform in $\phi$.

Two familiar limits follow from Equation~\ref{eq:forward}. If the boundary is steady in the co-rotating frame, the time derivative vanishes and propagation reduces to the HUX radial remapping, with a longitude shift set by solar rotation during the parcel travel time. If, in addition, interactions between streams are neglected so that each parcel keeps its speed, the characteristics become Parker-spiral-like paths and yield the ballistic, or time-of-flight, source mapping \cite{Nolte1973,MacNeil2022}. The ballistic mapping fails where it matters most for space weather. Fast wind overtakes slower wind launched earlier from neighboring longitudes, forming compression regions and CIRs, and during such intervals the speed at L1 reflects the history of several upstream streams rather than a single footpoint. A boundary estimate should therefore be obtained by training through a forward propagation, in which the travel time depends on the speed, rather than by assigning each L1 sample to one source location; Section~\ref{sec:solveradds} tests how much of the full nonlinear transport this requires.

Because Equations~\ref{eq:huxt_radial} and~\ref{eq:forward} are related by a change of coordinates, they describe the same physics and constrain the boundary in the same way. With $v_\varphi=0$, plasma reaching a spacecraft at time $T$ has travelled along the Sun--spacecraft line of that time, so the measured speed depends only on boundary values along one curve in the $(\phi,t)$ plane: at each departure time, the Carrington longitude that then lay on the arrival-time Sun--spacecraft line. Because Earth moves about $1^\circ$~d$^{-1}$ in its orbit, this differs by a few degrees from the longitude that faced the spacecraft at departure. For a single spacecraft at a fixed inertial azimuth, a one-dimensional solver along that line would therefore be equivalent. We use the co-rotating form for three reasons. It represents the whole boundary field on one fixed $(r,\phi,t)$ grid, so that the observer's orbital motion is handled exactly by the observation operator; one shared boundary can be compared with, or constrained by, spacecraft at different longitudes (this will be done in Part II); and smoothness across longitude can be imposed directly. In this paper only ACE is used for training.

The propagation is therefore the same reduced physics as HUXt \cite{Owens2020}, though the numerical scheme is slightly different (upwind versus a second-order finite-volume scheme). Here, it is solved in co-rotating form with fixed residual acceleration, and differentiated automatically. It shares the limitations of this model class: it omits pressure and magnetic forces and flow deflection, and stream interfaces can steepen into discontinuities. The contribution of this paper is the inversion, not the propagator.

\subsection{The inverse problem and its neural parameterization}
\label{sec:inverse}
Let the radial speed on the inner surface be an unknown time-dependent field, with boundary condition
\begin{equation}
 v_r(r_0,\phi,t)=u(\phi,t),\qquad r_0=21.5\,R_\odot.
 \label{eq:boundary}
\end{equation}
Let $\mathcal{P}$ denote the numerical propagation operator, which maps a prescribed boundary field $u(\phi,t)$ to the propagated radial-speed field $v_r(r,\phi,t)$ by solving the time-dependent transport model in Equation~\ref{eq:forward}, including the prescribed residual-acceleration profile described below. Let $\mathcal{H}_{\mathrm{ACE}}$ be the observation operator that interpolates this field to the measured heliocentric position and observation times of ACE. We define the forward operator as $\mathcal{F}_{\mathrm{ACE}}=\mathcal{H}_{\mathrm{ACE}}\circ\mathcal{P}$; hence, $\mathcal{F}_{\mathrm{ACE}}(u)$ is the modeled ACE speed time series corresponding to the boundary field $u$. With an unrestricted boundary, a generic variational inversion would solve
\begin{equation}
 u^*=\mathop{\operatorname{argmin}}_u
 \left\{\mathcal{L}\big(\mathcal{F}_{\mathrm{ACE}}(u),v_{\mathrm{ACE}}\big)
 +\mathcal{R}(u)\right\}.
 \label{eq:unrestricted}
\end{equation}
Here, $\mathcal{L}$ measures the discrepancy between the modeled ACE speed series $\mathcal{F}_{\mathrm{ACE}}(u)$ and the observed series $v_{\mathrm{ACE}}$; $\mathcal{R}(u)$ denotes regularization imposed on the boundary field.
This problem is underdetermined: the ACE time series near L1 does not sample the full longitude--time boundary. Nonlinear transport can also cause distinct upstream histories to produce similar downstream signals. Smoothness regularization alone is not expected to remove this ill-posedness: it selects the smoothest of the many boundary fields consistent with the ACE series, but adds no information about the boundary away from the sampled track or about which of several upstream histories actually occurred.

We therefore do not optimize each boundary value separately. Instead, a single neural mapping $g_\theta$, with parameters shared over longitude and time, transforms image-derived and auxiliary inputs $x(\phi,t)$ into the boundary field,
\begin{equation}
 u_\theta(\phi,t)=g_\theta\big(x(\phi,t)\big),
 \qquad
 \theta^*=\mathop{\operatorname{argmin}}_\theta
 \left\{\mathcal{L}\big(\mathcal{F}_{\mathrm{ACE}}(u_\theta),v_{\mathrm{ACE}}\big)
 +\mathcal{R}(u_\theta)\right\}.
 \label{eq:parameterized}
\end{equation}
Thus, the observational mapping is learned from discrepancies at ACE after nonlinear propagation, rather than fitted against an independently known boundary-speed target, which makes the approach a form of physics-informed self-supervised learning. Because Equation~\ref{eq:parameterized} optimizes the neural-network parameters $\theta$, standard machine-learning optimizers and automatic-differentiation software can train the model through the numerical propagation. In this paper $\mathcal{L}$ is the combined score $S$ of Section~\ref{sec:metric}, evaluated on ICME-free ACE hours, and $\mathcal{R}$ consists of the smoothness penalties of Equation~\ref{eq:loss}.

This is a constrained, amortized inverse reconstruction; it does not establish that the unobserved boundary field is unique or physically exact. In particular, the learned map from observations to boundary speed is optimal for the propagation model it is trained through, together with its residual acceleration, and is not a universal relation. This is a deliberate difference from the usual practice for empirical boundary relations such as WSA, whose parameters are calibrated against speeds at 1~AU with one, usually simple kinematic, mapping, and which are then used as the boundary of quite different propagation models, from HUX-type models to full MHD. A boundary relation calibrated with one propagation model carries any mismatch into the other, which is why WSA speeds are commonly adjusted before being propagated \cite{Owens2020,Riley2011}. In the inverse formulation the boundary and the propagation are fitted together, so the boundary is consistent with its propagation by construction; conversely, using it with a different propagation model would require retraining the map through that model.

Through $\mathcal{F}_{\mathrm{ACE}}$, a boundary value determines both the speed that reaches ACE and the time at which it arrives, so the gradient of $\mathcal{L}$ carries timing as well as amplitude information. Since ACE constrains the boundary only along its source track, the gradient of the loss reaches only the cells along that track. The rest of the boundary field is determined by the shared mapping $g_\theta$: the same function of the local image features is applied at every longitude and time, so what is learned along the track is applied elsewhere; the weak smoothness penalty adds only a little coupling. Away from the track, the reconstructed boundary is the network's prediction, not a fit to data.

\subsection{Discrete automatic differentiation and propagation configuration}
\label{sec:autodiff}
We differentiate the numerical forward map itself, yielding discrete gradients from ACE misfit through the propagated state to the image-to-boundary network. This avoids implementing an independent continuous adjoint and allows a common set of neural weights to be trained across many rotations. Automatic differentiation computes derivatives of the chosen discrete model; it neither eliminates inverse non-uniqueness nor removes model discrepancy \cite{Li2004,Fikl2016}.

\begin{figure*}[t]
\centering\includegraphics[width=\textwidth]{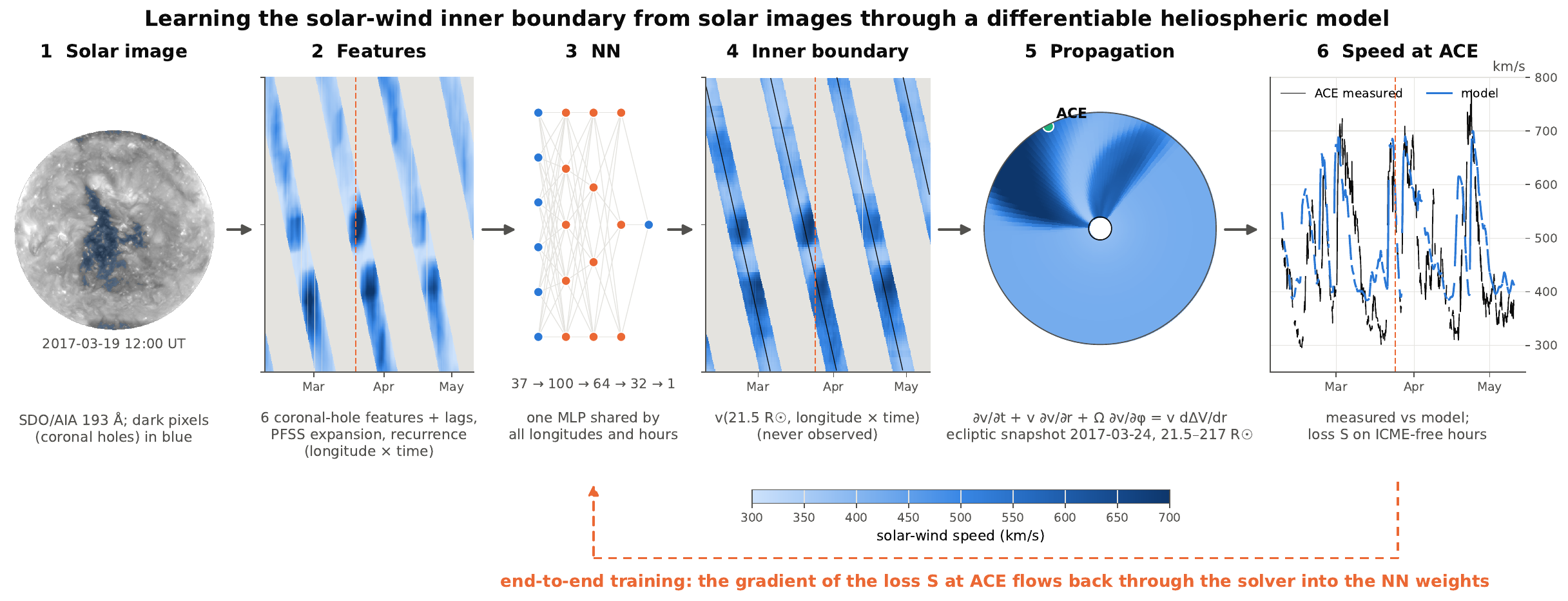}
\caption{Inverse-reconstruction workflow. Inputs defined per Carrington longitude and hour are mapped by one shared network (NN) to the speed at $21.5\,R_\odot$; the differentiable solver propagates this boundary to ACE, where the loss $S$ (Section~\ref{sec:metric}) is evaluated on measured, ICME-masked hours. The gradient of the loss flows back through the solver into the network weights. The inner-boundary speed itself is never observed. Only the cells along the sub-Earth track (solid black lines in panel 4) are constrained by ACE; elsewhere the boundary is the network's prediction from the same features.}
\label{fig:pipeline}
\end{figure*}

The numerical domain extends from $21.5$ to $217\,R_\odot$, with 64 radial grid points ($\Delta r=3.10\,R_\odot$), 128 periodic Carrington-longitude cells ($\Delta\phi=2.8125^\circ$) and a time step of 1800~s.Each solver run covers one chunk of about 86~days, preceded by a 150-h warm-up (Section~\ref{sec:train}), so a run spans about 92~days, or some 4,400 time steps; the development interval is covered by 40 such runs. Model outputs are interpolated in time to ACE's actual heliocentric distance and longitude (the outer edge exceeds the spacecraft's maximum distance near 1~AU). We prescribe a residual-acceleration profile with the exponential radial dependence used in HUX and HUXt \cite{Riley2011,Owens2020}, but with a fixed amplitude in this study:
\begin{equation}
 \Delta V(r)=V_{\mathrm{ref}}\left[1-\exp\left(-\frac{r-r_0}{L}\right)\right],
 \qquad V_{\mathrm{ref}}=40\ \mathrm{km\,s^{-1}},\quad L=50R_\odot.
 \label{eq:acceleration}
\end{equation}
In HUX and HUXt the amplitude is proportional to the boundary speed ($\alpha u$, with $\alpha\approx0.15$); here it is the same for all speeds.
It enters the transport as a source term, so that the solved equation is
\begin{equation}
\frac{\partial v_r}{\partial t}+v_r\frac{\partial v_r}{\partial r}
-\Omega\frac{\partial v_r}{\partial\phi}=v_r\frac{\mathrm{d}\Delta V}{\mathrm{d}r};
\label{eq:forward_accel}
\end{equation}
for steady radial flow this adds $\Delta V(r)$ to the speed of every parcel between $r_0$ and $r$. Although $V_{\mathrm{ref}}$ and $L$ could in principle be learned, they are held fixed in the final experiments. Their separate values cannot be reliably inferred from the ACE speeds: re-scoring a trained model over a grid of values gives a nearly flat valley of configurations with similar net acceleration, and the net offset trades off against the level of the boundary speed. 
The numerical discretization is described in \ref{app:numerics} and verified against exact solutions in Figure~\ref{fig:verification}. Figure~\ref{fig:pipeline} summarizes the end-to-end computational graph.

\section{Data}
\label{sec:observations}
\subsection{Solar wind speed at L1}
The primary target is the hourly proton bulk speed measured by the Solar Wind Electron, Proton, and Alpha Monitor (SWEPAM) on the Advanced Composition Explorer (ACE) \cite{Stone1998,McComas1998}, taken from \url{https://sohoftp.nascom.nasa.gov/sdb/goes/ace}. Speeds outside 0--3000~km~s$^{-1}$ are removed. Because this paper addresses ambient solar wind rather than transient CME ejecta, we exclude intervals associated with interplanetary CMEs (ICMEs), from the event start through the end of the magnetic obstacle as listed for Wind at L1 in HELIO4CAST ICMECAT v2.3 \cite{Moestl2025ICMECAT}.
We define our development interval used for training, validation and test  from 1 January 2011 to 1 June 2020 at 12:00 UTC (82,548 hours), covering the rise, maximum, decline and minimum of solar cycle 24. Of these hours, 4,380 (5.3\%), belonging to 139 ICMEs, are masked. A further 7,817 hours lack a valid ACE speed, which leaves 70,351 measured hours. Data after 1 June 2020 were not used at any stage of development and form the independent test of Part~II. 

\subsection{EUV solar images}
Solar imaging comes from the Atmospheric Imaging Assembly (AIA; \citeA{Lemen2012}) on the Solar Dynamics Observatory (SDO; \citeA{Pesnell2012}) synoptic Level-1.5 product (1024 by 1024 pixels; one usable image per hour). We retain frames with flags $\mathrm{QUALITY}=0$ and $\mathrm{EXPTIME}\geq0.1$~s; the few frames shortened by automatic exposure control during large flares were replaced (\ref{app:data}).
We normalize each frame by its exposure time (DNs$^{-1}$) and correct for the time-dependent sensitivity loss of the 193\AA{} channel with the degradation factor of aiapy \cite{Barnes2020aiapy} (version 0.10.2; JSOC correction table \texttt{aia.response}, calibration version~10).
Pixels are mapped to heliocentric ecliptic latitude $\beta$ and to their longitude offset from the central meridian, and pixels beyond $\mu=\cos78^\circ$ from disk center are excluded.

\subsection{HMI magnetograms}
Photospheric magnetic quantities for the feature screening are computed from the \texttt{hmi.M\_720s} line-of-sight magnetograms of the Helioseismic and Magnetic Imager (HMI; \citeA{Scherrer2012}) on SDO, one per hour (the 720-s magnetogram at the full hour), with the radial field estimated as $B_r=B_{\mathrm{los}}/\mu$, where $\mu$ is the cosine of the heliocentric angle; pixels with $\mu<\cos65^\circ$ are excluded.
They enter only the screening (Section~\ref{sec:screening}) and a baseline regression introduced in Section~\ref{sec:train}; the reference model does not use them.

\subsection{Photospheric magnetic field and PFSS expansion factor}
\label{sec:pfss}
The coronal magnetic field is represented by potential-field source-surface (PFSS) extrapolations \cite{Altschuler1969,Schatten1969} of GONG synoptic magnetograms \cite{Harvey1996}. We use the zero-point-corrected, hourly updated GONG synoptic product (\texttt{mrzqs}), resampled to $360\times180$ pixels in Carrington longitude and sine latitude. Each map is extrapolated with the finite-difference PFSS solver of sunkit-magex (formerly pfsspy; \citeA{Stansby2020pfsspy}), with the source surface at $R_{\mathrm{ss}}=2.5\,R_\odot$ and 30 radial grid points.

From each extrapolation we trace one field line downward from just below the source surface at each of the 128 model longitudes, $\phi_j=j\cdot360^\circ/128$, starting at the heliographic latitude of Earth, $B_0$, at the map time. For each open line we record the expansion factor
\begin{equation}
 f_s=\left(\frac{R_\odot}{R_{\mathrm{ss}}}\right)^2\frac{B(R_\odot)}{B(R_{\mathrm{ss}})},
 \label{eq:fs}
\end{equation}
with $B$ the field strength at the two ends of the line \cite{WangSheeley1990}, and use $\log_{10}f_s$ as the model input; longitudes whose line is closed or fails to trace take the value of the nearest valid longitude.
We use the expansion factor alone; the distance of the footpoint from the nearest coronal-hole boundary, the second ingredient of WSA, is represented instead by the AIA hole-boundary feature (Section~\ref{sec:features}).

The input at each hour comes from the most recent GONG map at or before that hour, so it uses no later magnetogram information; like the other inputs, it is standardized with the statistics of the training batches. A version using only the first map of each day gave the same skill (Section~\ref{sec:ablations}), so the hourly cadence is not essential. Processing statistics are given in \ref{app:data}.

\section{Feature extraction and ranking}
\subsection{Physically motivated feature families}
\label{sec:features}
Features are constructed from individual pixels in a heliocentric ecliptic coordinate system, using their latitude $\beta$, longitude offset from the observing meridian and valid-disk masks. A 193~\AA{} pixel is classified as dark when its intensity falls below a fraction $f$ of that frame's disk median. The candidates fall into five families (Table~\ref{tab:allfeatures}): (A) dark area in full-height central strips of different longitudinal widths and thresholds; (B) the proximity of dark regions to the ecliptic, near which the wind that reaches Earth originates, measured in three ways: the dark area near the ecliptic, how close the nearest coronal hole comes to it, and whether the ecliptic lies inside a hole; (C) latitude profiles of the dark fraction and of the intensity; (D) the darkness of the dark pixels, through their intensity distribution and integrated depth; and (H) HMI field strength, polarity-inversion-line distance and flux imbalance. %Together they test whether the wind is better predicted by the area of the coronal-hole source (A), by its likely connection to the ecliptic (B), by its structure in latitude (C), or by how dark the coronal-hole pixels are rather than only whether they are dark (D). 

We consider 42 candidate features in all, defined in Table~\ref{tab:allfeatures}. After the ranking and selection of Section~\ref{sec:screening}, six are retained; together with their recent history, they form the image inputs of the reference model, which we call channels (Table~\ref{tab:channels}).

All intensities are relative to the median $I_{\mathrm{med}}$ of the pixels within $0.98\,R_\odot$ of disk centre in the same frame. A dark-area feature is the percentage of pixels in its region with $I<f\,I_{\mathrm{med}}$, counted among the pixels of the region with $\mu\geq\cos78^\circ$; pixels are not weighted by the solar surface they cover, so the feature is the dark fraction of the region as seen from Earth. The darkness feature (channel 3) is the mean $\log_{10}(I/I_{\mathrm{med}})$ of the dark pixels ($f=0.35$) in its region, and the
hole-boundary distance (channel 4) is measured on the grid of $0.5^\circ$ cells in $(\beta,\Delta\mathrm{lon})$ on which the coronal-hole mask is defined.

\begin{figure*}[t]
\centering\includegraphics[width=\textwidth]{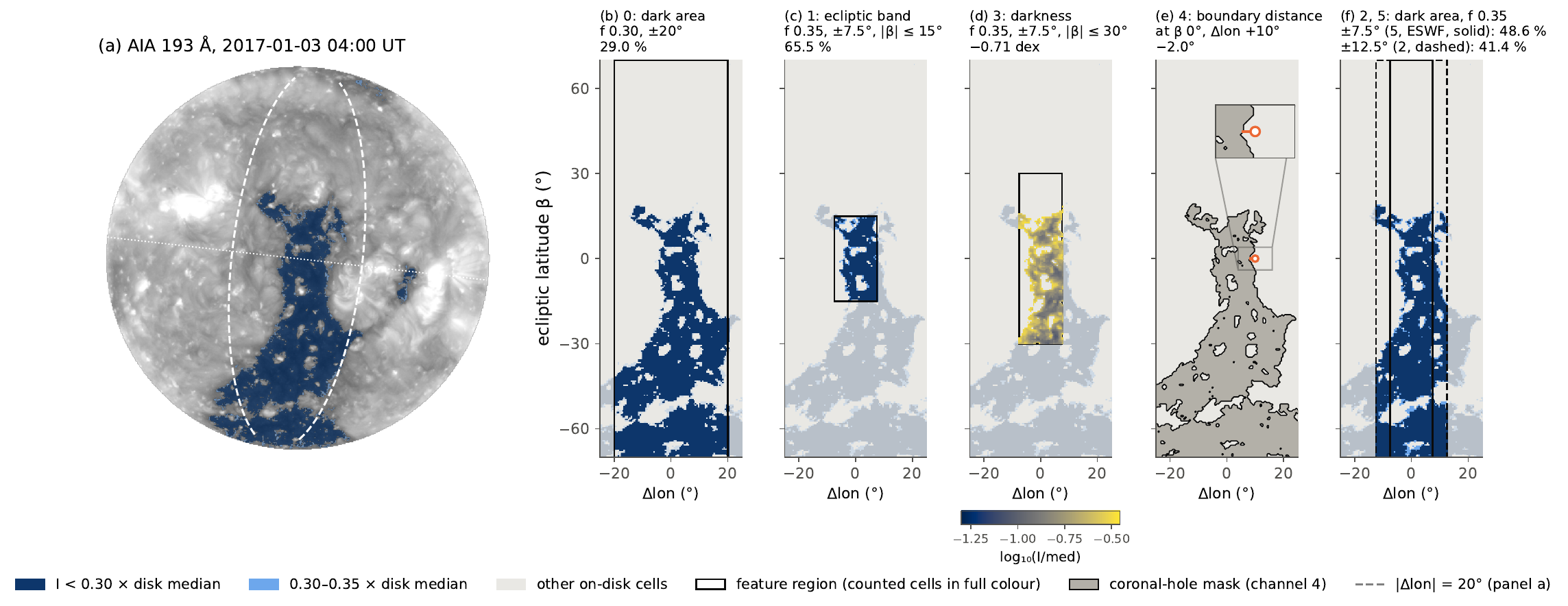}
\caption{Example of six AIA 193~\AA{} features of the model on one frame (3 January 2017, 04:00~UT).
(a) $\log_{10}(I/I_{\mathrm{med}})$ with the two dark levels used by the features: $I<0.30\,I_{\mathrm{med}}$ (dark blue)
and $0.30$-$0.35\,I_{\mathrm{med}}$ (light blue); dashed: $|\Delta\mathrm{lon}|=20^\circ$; dotted: the ecliptic.
(b)--(f) The channels of Table~\ref{tab:channels}, in the $0.5^\circ$ cells of ecliptic latitude $\beta$ and longitude
$\Delta\mathrm{lon}$ from the central meridian in which the features are computed; a cell is shown dark when most of its
pixels are. The outlined region is the feature's, and the cells it counts are in full colour. Channels 2 and 5 differ only
in the strip width and share panel (f) (dashed: $\pm12.5^\circ$; solid: $\pm7.5^\circ$). Channel 3 is the mean
$\log_{10}(I/I_{\mathrm{med}})$ of the dark pixels in its region, shown in colour. Channel 4 is the signed distance from
($\beta=0^\circ$, $\Delta\mathrm{lon}=+10^\circ$) to the nearest edge of the coronal-hole mask (grey: connected cells whose
dark fraction at $f=0.35$ is at least 0.5), positive inside a hole; inset: the point and the distance. The value of each
feature for this frame is given above its panel.}
\label{fig:features}
\end{figure*}

Figure~\ref{fig:features} depicts graphically, on a single frame, the region of each selected feature, the pixels it counts and its value.

\begin{figure*}[t]
\centering\includegraphics[width=\textwidth]{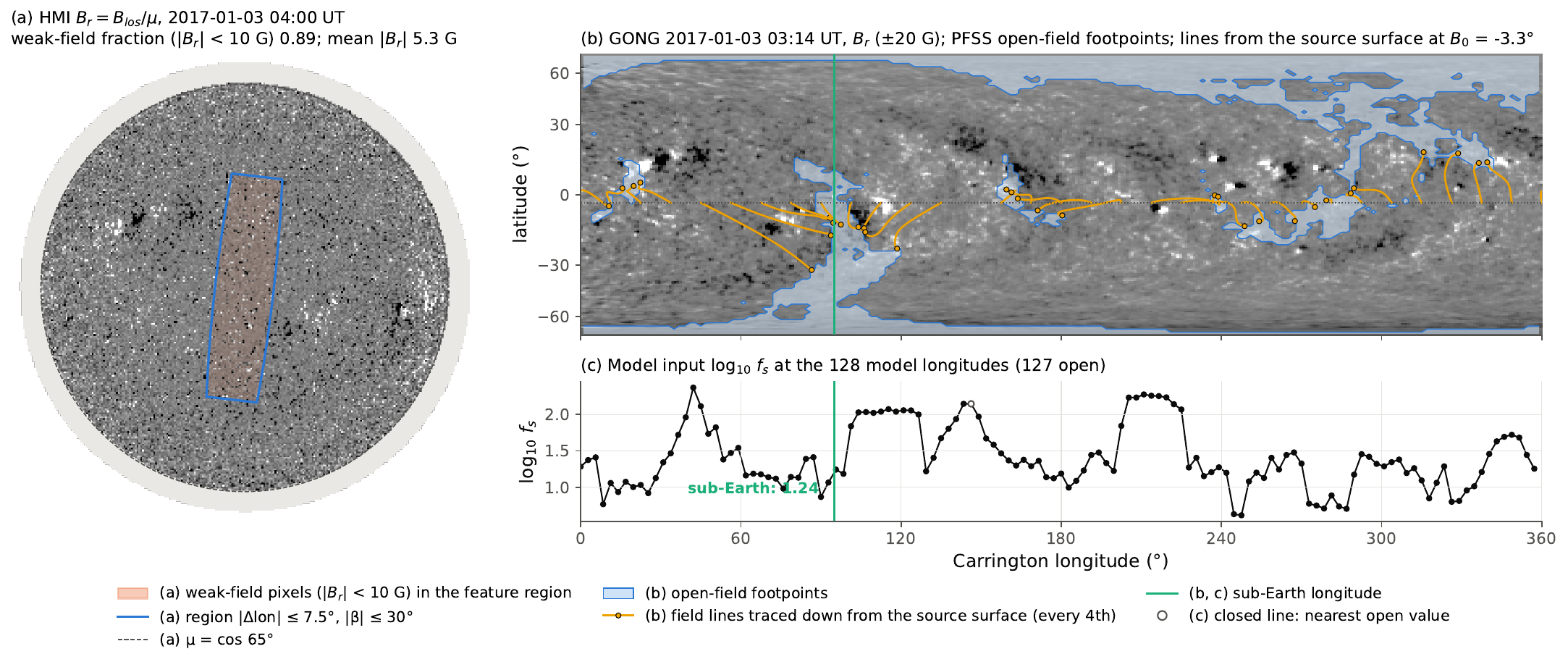}
\caption{The magnetic inputs at the hour of Figure~\ref{fig:features}. (a) HMI radial field
$B_r=B_{\mathrm{los}}/\mu$ ($\mu\geq\cos65^\circ$), with the region of the two HMI features of the linear comparator
($|\Delta\mathrm{lon}|\leq7.5^\circ$, $|\beta|\leq30^\circ$) and its weak-field pixels ($|B_r|<10$~G). (b) The GONG
synoptic map used by the reference model at this hour (the most recent before it, 03:14~UT), $B_r$ in Carrington longitude and
sine latitude, with the photospheric footpoints of open field in the PFSS extrapolation (shaded) and the field lines traced
down from the source surface at Earth's heliographic latitude $B_0$ for every fourth model longitude, with their
footpoints. (c) The model input $\log_{10}f_s$ at the 128 model longitudes; the open circle marks a longitude whose line is
closed and which takes the value of the nearest open longitude. Green: the sub-Earth longitude.}
\label{fig:magnetic_inputs}
\end{figure*}

Figure~\ref{fig:magnetic_inputs}a shows the radial field and the region of the two HMI features that the screening favoured for detecting fast wind (Section~\ref{sec:screening}), the fraction of weak-field pixels and the mean field strength; in this frame 89\% of its pixels have $|B_r|<10$~G. The end-to-end tests later showed that they do not improve the model, so the reference model does not use them; they enter only a baseline regression introduced in Section~\ref{sec:train}.

Figure~\ref{fig:magnetic_inputs}b,c shows the extrapolation and the resulting input for the same hour. At the sub-Earth longitude the open-field region reaches from the equator to the south pole,
the coronal hole seen in the AIA image, and $\log_{10}f_s=1.24$, against about 2 at the neighbouring longitudes
105--$125^\circ$.

\subsection{Feature ranking and selection}
\label{sec:screening}
To rank candidate features, we evaluate their association with ACE speed four days after image acquisition. This fixed-lead target follows common practice in empirical solar-wind forecasting, but differs from the inverse problem, which has no single fixed lead time. In the inverse model, arrival time emerges from propagation. We pair each image with the ACE speed 96~h later. We selected the 96-h lag as the one that maximizes the Pearson correlation association of the best correlated feature (mean within-batch correlation 0.48); tracing each hour back ballistically at its own observed speed gives a lower correlation (0.43) and the same ranking.
The development interval is divided into five contiguous batches of about 688 days each, which are also the cross-validation folds of the model (Section~\ref{sec:train}). We paired each image time with the ACE speed four days later, evaluated all 42 candidates (AIA, HMI and reference predictors) at that time, and kept the 49,183 pairs for which every candidate could be computed, so that all candidates are compared on the same hours. Each candidate is scored by its Pearson correlation with the speed, computed separately within each batch and averaged over the five, so that a feature cannot appear predictive merely because both it and the speed follow the solar cycle; uncertainties come from a block bootstrap over whole solar rotations (\ref{app:screening}, which also gives the ranking of all candidates).
The ranking and selection used the whole development interval, including the batches that later serve as test folds, so the scores reported later may be slightly better than they would be on unseen data. The effect is small: along the backward elimination described below, every subset from five to 41 features gives a linear regression with an out-of-sample RMSE between 75.6 and 77.1~km~s$^{-1}$, so choosing among them can have gained little from looking at these batches. Nevertheless, it is one reason the independent test is reserved for Part~II.

The best single predictor is the dark area with threshold $f=0.30$ in a strip of half-width $20^\circ$ about the central meridian, with a mean correlation of 0.475 (95\% interval 0.429--0.515). It is a variant of the predictor of the Empirical Solar Wind Forecast (ESWF), the fractional coronal-hole area within a meridional slice of $\pm7.5^\circ$, which ESWF relates linearly to the speed at 1~AU about four days later (average delay $4.02\pm0.5$~d; \citeA{Rotter2012,Rotter2015}). Our version uses a wider strip and a darker threshold; our implementation of the ESWF area ($f=0.35$, half-width $7.5^\circ$) reaches only a correlation 0.42. These correlations are computed over all hours and are not directly comparable with the values reported for ESWF, which refer to the peak speeds of high-speed streams ($cc\approx0.60$ for 156 streams in 2011--2013; \citeA{Rotter2015}).

Many candidates carry overlapping information, so we selected a compact set by backward elimination: a linear regression of the speed on all standardized candidates is fitted, the candidate that adds least is removed, and the procedure is repeated, scoring each step on held-out batches; the same procedure was repeated with a logistic classifier of fast wind (\ref{app:screening} for full details). In the regression setting, a linear regression on only five features is as accurate as the best subset found (out-of-sample RMSE 75.7 against 75.6~km~s$^{-1}$ with 15 features), so we kept these five: the dark area at $f=0.30$ within $\pm20^\circ$ and at $f=0.35$ within $\pm12.5^\circ$, the dark fraction within $15^\circ$ of the ecliptic, the darkness of the dark pixels, and the distance to the nearest hole boundary $10^\circ$ west of the meridian. In the classification setting, the classifier of fast wind favoured three more: the ESWF area and two HMI quantities of the quiet field near the ecliptic. Restoring the two HMI features in the end-to-end model did not improve its validation score ($S$ $+0.4\pm0.6$~km~s$^{-1}$; Section~\ref{sec:ablations}), so the final selection is six AIA features (Table~\ref{tab:channels}). With their values 12, 24, 48 and 72~h earlier, the PFSS expansion factor, the back-mapped recurrence (the speed observed one solar rotation earlier at the same source longitude) and flags for missing values, the model has 37 inputs (Section~\ref{sec:model}).

\subsection{Off-meridian observability}
\label{sec:offmeridian}
Each feature is computed every hour for each Carrington longitude within $60^\circ$ of disk center. Toward the limb the same coronal hole can look different, because of foreshortening, limb brightening and bright structures in front of it: a feature measured $55^\circ$ from the central meridian correlates with its value at meridian crossing only at about 0.7 for the area features and 0.36 for the hole-boundary distance (\ref{app:data}). We remove the systematic part of this change by rescaling each feature as a function of its distance from the meridian, with factors fitted on the training batches, and apply the same feature-to-speed map at all distances; the rescaling cannot recover information hidden near the limb. Off-meridian views also enter through the lagged inputs: 12, 24, 48 and 72~h before a longitude reaches the sub-Earth point, it lies about $7^\circ$, $13^\circ$, $26^\circ$ and $40^\circ$ east of the central meridian.

\section{Input ambiguity and noise analysis}
\label{sec:degeneracy}
If the inputs determined the speed, situations with similar inputs would be followed by similar speeds. We test this in two ways, both of which pair every measured ACE hour with the inputs of one boundary cell. A look-alike analysis asks, independently of any trained model, how often the inputs that preceded fast wind resemble those that preceded slow wind. A noise estimate quantifies the part of the variance that no model using these inputs can explain. It is the one of the three terms of the bias--variance--noise decomposition of the squared error; the other two, the variance due to training and the bias, depend on the trained model and are estimated in Section~\ref{sec:budget}.

\subsection{Look-alikes}
The look-alike analysis asks, for each fast-wind hour, whether situations that looked like it on the Sun were followed by fast wind as well. Each ACE hour is paired with the inputs of the cell that crossed the central meridian 96~h earlier, at time $t_{\mathrm{src}}$, the alignment used in the feature screening (Section~\ref{sec:screening}). We describe each hour by 26 values that summarize the model's inputs: the six selected features at $t_{\mathrm{src}}$ and 24, 48 and 72~h earlier, the expansion factor and the back-mapped recurrence. The 12-h lag, which differs little from the current value, and the binary availability flags are left out.

For each fast-wind hour (speed of at least 550~km~s$^{-1}$; 2,883 hours) we find the 50 hours whose 26 values are closest, its nearest neighbours. To avoid matching an hour with its own stream or its own phase of the solar cycle, neighbours must lie in another development batch and at least 13~days away. A neighbour counts as slow if the speed that followed it was at most 400~km~s$^{-1}$.

Consecutive hours of the same stream have nearly the same inputs and outcomes, so we count streams rather than hours. Fast hours less than 12~h apart form one stream, which gives 225 streams; this definition serves only this analysis and differs from the 187 catalogued events of Section~\ref{sec:cases}. For each hour of a stream we compute the fraction of its 50 neighbours that were followed by slow wind, and average this fraction over the stream's hours: we call the result the stream's slow-neighbour fraction. A stream is ambiguous when its slow-neighbour fraction exceeds 50\%, that is, when most of the situations that resembled it were followed by slow wind.

As a reference, we repeat the count with the neighbours replaced by randomly chosen hours, which carry no information about the stream. Each stream's slow-neighbour fraction is then close to 49\%, the fraction of all hours followed by slow wind.

The analysis shows that 16\% of the fast streams are ambiguous (Table~\ref{tab:lookalikes}): for about one fast stream in six, the available information resembles situations that were followed by slow wind, so a point prediction from these inputs is likely to miss it. Using only the six image features at $t_{\mathrm{src}}$, without their history, the expansion factor or the recurrence, gives the same fraction, but the additional inputs make the typical stream less ambiguous: the fraction of streams with more than a quarter of slow neighbours falls from 58\% to 41\%, and the median slow-neighbour fraction from 29\% to 21\% (mean 27\%). 

\begin{figure*}[t]
\centering\includegraphics[width=\textwidth]{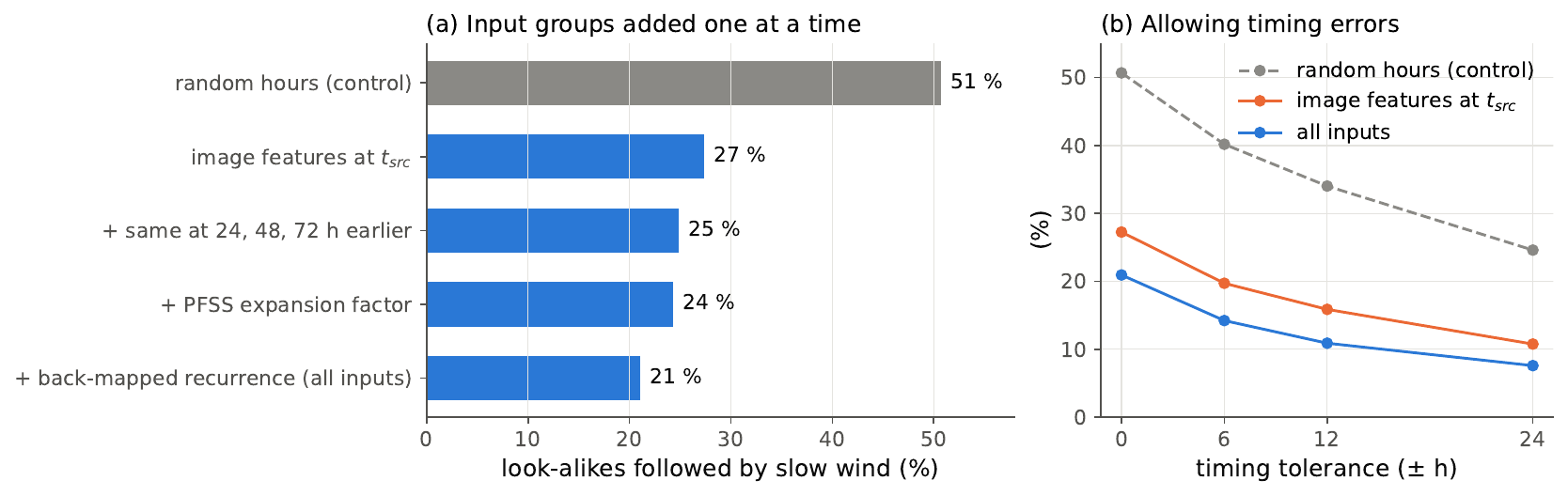}
\caption{Ambiguity of the inputs. (a) Fraction of fast-wind hours ($\geq550$~km~s$^{-1}$) whose 50 nearest look-alikes in input space were followed by slow wind ($\leq400$~km~s$^{-1}$), as input groups are added; the last bar adds the speed observed one rotation before the target hour, which the model does not receive; grey: random hours. (b) The same fraction when the outcome may be shifted in time by up to the given tolerance. Inputs are aligned with the fixed 96-h lag (Section~\ref{sec:screening}); look-alikes come from other development batches and are at least 13~days away.}
\label{fig:degeneracy}
\end{figure*}

Figure~\ref{fig:degeneracy}a shows how the hourly fraction of slow-wind look-alikes falls as input groups are added: from 49\% for random hours to 27\% with the six image features at $t_{\mathrm{src}}$, 25\% with their history, 24\% with the expansion factor and 21\% with the back-mapped recurrence. The ambiguity is not only a matter of timing.
Figure~\ref{fig:degeneracy}b repeats the count with a stricter criterion, under which a look-alike of a fast hour counts as slow only if its speed stays at or below 400~km~s$^{-1}$ throughout a window of $\pm$ the given tolerance. For the best input set tested (the model's information and the speed one rotation before the target hour), 8\% of the look-alikes
remain slow at $\pm24$~h, against 24\% for random hours: relative to inputs without information, the fraction falls from 0.42 to 0.32 of the random value, so timing errors explain part of the ambiguity but not all of it.

\subsection{Noise estimate}
The trained models connect each hour to the boundary through a speed-dependent propagation, so here each hour is paired with the cell it came from in the reference model's out-of-fold solution, traced backward along the characteristics of that solution from ACE to $21.5\,R_\odot$. For each hour we take its ten nearest neighbours in the space of the reference model's 32 continuous inputs (the 37 inputs without the five availability flags). As in the look-alike analysis, neighbours must lie in another development batch and at least 13~days away. Because some lagged images are missing, neighbours must also have the same lagged images available as the hour itself, so that the distance between them is computed over the same inputs.
The Gamma test \cite{Stefansson1997} asks how much the speeds of two hours still differ when their inputs become identical. For each $k=1,\dots,10$ it computes two averages over all hours $i$, using the $k$-th nearest neighbour $n(i,k)$ of each hour:
\begin{equation}
 \delta_k=\frac{1}{N}\sum_i\left|\mathbf{x}_{n(i,k)}-\mathbf{x}_i\right|^2,
 \qquad
 \gamma_k=\frac{1}{2N}\sum_i\left(y_{n(i,k)}-y_i\right)^2,
\end{equation}
where $\mathbf{x}$ are the standardized inputs and $y$ the observed speed: $\delta_k$ measures how far apart the inputs of $k$-th neighbours are, and $\gamma_k$ how different their speeds are. Neighbours of higher order are farther apart in input space and, if the speed depends on the inputs, differ more in speed. A straight line $\gamma_k=\Gamma+A\,\delta_k$ is fitted through the ten points $(\delta_k,\gamma_k)$, and its intercept $\Gamma$, the speed difference extrapolated to identical inputs, estimates the noise variance $\mathrm{E}[\mathrm{Var}(Y\mid X)]$.

Over the 65,047 hours with a traced source and complete inputs, the noise is 67.6~km~s$^{-1}$ RMS (95\% interval from resampled 27-day blocks: 64.0--71.2~km~s$^{-1}$), against a standard deviation of 89.5~km~s$^{-1}$ of the observed speed on the same hours. About 57\% of the variance of the hourly speed is therefore not determined by these inputs, which would limit a model that uses them to a correlation of about 0.65; the reference model reaches 0.58 (Section~\ref{sec:accuracy}). The noise rises from 59~km~s$^{-1}$ where the inputs point to slow wind to 79--88~km~s$^{-1}$ where they point to wind faster than about 450~km~s$^{-1}$. The estimate is conditional on the inputs, on the reference model's propagation and on the extrapolation to zero distance. %Two other alignments bracket it: matching each hour ballistically at its own observed speed gives 60.5~km~s$^{-1}$, which is optimistic because the alignment then uses the speed to be explained; a fixed 96-h lag gives 69.6~km~s$^{-1}$, but this exceeds the error of a trained model in two speed regimes, because one cell at a fixed lag under-represents the information that the models use.

The two analyses lead to the same conclusion. The inputs used here were chosen from the standard physical predictors of the solar-wind speed, coronal-hole properties, the flux-tube expansion factor and the speed one rotation earlier, by an objective ranking, and yet they leave a large part of the speed undetermined: about one fast stream in six resembles situations that were followed by slow wind, and more than half of the hourly variance is noise given these inputs. A more flexible model, such as a deeper network, cannot remove this noise; only better or additional inputs can substantially improve the reconstruction. Both estimates depend on the chosen inputs, distance and alignment, so they describe these inputs rather than set a universal bound on what any reconstruction can achieve.

\begin{table}[t]
\caption{Slow-wind outcomes among the 50 nearest neighbours of fast-wind hours ($\geq550$~km~s$^{-1}$), over 225 fast-wind streams. A stream's slow-neighbour fraction is the mean, over its hours, of the fraction of neighbours followed by slow wind ($\leq400$~km~s$^{-1}$); a stream is ambiguous when its slow-neighbour fraction exceeds 50\%.}
\label{tab:lookalikes}\centering
\begin{tabular}{lcccc}\hline
 & Ambiguous & \multicolumn{2}{c}{Slow-neighbour fraction of streams} & Hourly\\
Inputs & streams & $>$\,25\% & mean (median) & fraction\\\hline
Model inputs (26 values) & \textbf{16\%} & \textbf{41\%} & \textbf{27\% (21\%)} & \textbf{21\%}\\
Six image features at $t_{\mathrm{src}}$ only & \textbf{16\%} & 58\% & 32\% (29\%) & 27\%\\
Random hours (no information) & 40\% & 100\% & 50\% (50\%) & 49\%\\\hline
\end{tabular}
\end{table}

\section{End-to-end learning methodology}
\label{sec:model}
\subsection{Shared neural boundary mapping}
The configuration used throughout this paper, which we call the reference model, maps the inputs of each boundary cell to its speed with a multilayer perceptron of widths $37\to100\to64\to32\to1$ and GELU activations (12,377 parameters), shared by all longitudes and times; a sigmoid maps its output to the interval 150--800~km~s$^{-1}$. The 37 inputs (Table~\ref{tab:channels}) are the six selected 193~\AA{} features of the cell's Carrington longitude; the same features 12, 24, 48 and 72~h earlier, each lag with a flag marking a missing image; $\log_{10}f_s$ from the most recent GONG map; and a back-mapped recurrence speed with its validity flag. Off-meridian and lagged values are rescaled to their central-meridian equivalents (Section~\ref{sec:offmeridian}), and all inputs are standardized with the statistics of the training batches.

The back-mapped recurrence is the mean ACE speed whose ballistic source, one rotation earlier, was the cell's longitude. It uses source times within $\pm5$~d of one rotation (27.27~d) before the boundary time $t$, so the latest source time is $t-22.3$~d; since even the slowest observed wind (234~km~s$^{-1}$) reaches ACE within 6.7~d, every observation used is at least 15.7~d older than $t$. The input is therefore causal and could be computed in real time; this minimum age follows from the window and the slowest observed transit, and was not tuned. The ACE speeds used are the same ICME-masked series as the training target, so CME ejecta do not enter the input; when no valid hour remains in a cell's window, its validity flag is set to 0. This input carries the same rotation-old information that assimilation schemes such as BRaVDA use to update the boundary (Section~\ref{sec:intro}), but it enters the learned boundary map as an input rather than through a variational update.

%Cells without a valid image are assigned the training-median speed as a fixed fallback that carries no gradient. The output range (up to 840~km~s$^{-1}$ at ACE with the residual acceleration) covers all but 7 of the 70,351 development hours (maximum 853~km~s$^{-1}$), and no boundary value of the reference model comes within 1\% of either bound, so the sigmoid never saturates. A second network applied after the propagation, as a residual correction at ACE, was tested in preliminary configurations and added no skill once the recurrence input was included; the HMI inputs were dropped for the reason given in Section~\ref{sec:screening}. The remaining choices follow the ablations of Section~\ref{sec:ablations}.

\begin{table}
\caption{Inputs of the reference model for one boundary cell (Carrington longitude, hour $t$). Channels 0--5 are the six AIA 193~\AA\ features selected in Section~\ref{sec:screening}, rescaled to their central-meridian equivalent; the full definitions and the screening statistics of all candidates are in Table~\ref{tab:allfeatures}. All inputs are standardized with training-batch statistics.}
\label{tab:channels}\centering\small
\begin{tabular}{lp{0.72\linewidth}r}\hline
Channel & Input & Values\\\hline
0 & Dark area ($f=0.30$), strip of half-width $20^\circ$ centred on the meridian & 1\\
1 & Dark fraction within $|\beta|\leq15^\circ$, central strip & 1\\
2 & Dark area ($f=0.35$), strip of half-width $12.5^\circ$ centred on the meridian & 1\\
3 & Mean $\log_{10}(I/I_{\mathrm{med}})$ of the dark pixels, central strip, $|\beta|\leq30^\circ$ & 1\\
4 & Signed distance from the ecliptic to the nearest hole boundary, 10$^\circ$ west of the meridian (positive inside a hole) & 1\\
5 & Dark area ($f=0.35$), half-width $7.5^\circ$ (ESWF coronal-hole area) & 1\\
6--33 & Channels 0--5 at $t-12$, $t-24$, $t-48$ and $t-72$~h, each lag with a flag marking a missing image & 28\\
34 & $\log_{10}$ of the PFSS expansion factor $f_s$ at the cell (hourly GONG map) & 1\\
35--36 & Back-mapped recurrence: mean ACE speed traced back to this longitude one rotation earlier, and its validity flag & 2\\
\hline
 & Total & 37\\\hline
\end{tabular}
\end{table}

\subsection{TIMSE, reconstruction scores and regularization}
\label{sec:metric}
A model trained on the mean squared error predicts, for given inputs, the average of the speeds that may follow them. When similar inputs can be followed by either fast or slow wind (Section~\ref{sec:degeneracy}), this average lies between the two, so relatively rare high-speed streams are predicted with too little amplitude. To weight errors across the whole observed speed distribution, we introduce the threshold-integrated mean squared error (TIMSE): the mean squared error over the hours whose observed speed exceeds a threshold, integrated over all thresholds. Errors at fast hours enter the integral for many thresholds and therefore weigh more, without the choice of a single event threshold. TIMSE is related to threshold-weighted scores for probabilistic forecasts \cite{Gneiting2011}, applied here to point predictions.
For observed speeds $y_i$ and predictions $\hat y_i$, we use as thresholds the 0th to 100th percentiles of the observed speeds, $q_k$ for $k=0,\dots,K$ with $K=100$ (linear interpolation between order statistics), so that $q_0=y_{\min}$ and $q_K=y_{\max}$. For each threshold we count the hours with an observed speed at least as large and compute the mean squared error over them,
\begin{equation}
 N_k=\#\{i:y_i\geq q_k\},\qquad
 M_k=\frac{1}{N_k}\sum_{i:y_i\geq q_k}(y_i-\hat y_i)^2.
 \label{eq:tailmse}
\end{equation}
$M_k$ is the error on the tail of the speed distribution above $q_k$: at the lowest threshold it is the ordinary mean squared error over all hours, and as the threshold rises it is computed over fewer and faster hours. TIMSE integrates this tail error over all thresholds, normalized by the observed speed range:
\begin{equation}
 \operatorname{TIMSE}_N=\frac{1}{y_{\max}-y_{\min}}\sum_{k=0}^{K-1}(q_{k+1}-q_k)\,M_k.
 \label{eq:timse_n}
\end{equation}
Since the interval widths sum to the range, $\operatorname{TIMSE}_N$ is a weighted average of the tail errors and has units of squared speed. The percentiles are closely spaced in slow wind, where hours are many, and widely spaced in the sparse fast tail, so the tail errors of fast wind receive most of the weight.

Exchanging the two sums shows what this does to each hour:
\begin{equation}
 \operatorname{TIMSE}_N=\sum_i w_i(y_i-\hat y_i)^2,
 \qquad
 w_i=\frac{1}{y_{\max}-y_{\min}}\sum_{k<K:\,q_k\leq y_i}\frac{q_{k+1}-q_k}{N_k}.
 \label{eq:timse}
\end{equation}
The squared error of each hour receives a weight $w_i$ that sums over all thresholds below its speed. A slow hour lies above few thresholds and receives a small weight; a fast hour lies above many, each shared with fewer hours, and receives a large one. TIMSE therefore weights errors increasingly with the observed speed, without singling out one event threshold.

TIMSE alone would neglect the slow wind that makes up most hours, so the score used for model selection and training combines it with the ordinary mean squared error through their geometric mean,
\begin{equation}
 S=\left(\mathrm{MSE}\cdot\operatorname{TIMSE}_N\right)^{1/4},
 \label{eq:score}
\end{equation}
so that $S^2$ is the geometric mean of the two errors and $S$, like the RMSE, is a speed. The geometric mean makes the score sensitive to relative changes in either error: a model cannot compensate a large tail error with a small overall error, or vice versa, as easily as with an arithmetic mean.
During training the percentile grid is computed once from all training-batch observations, so the weights do not change between optimization steps; reported scores use the grid of the scored observations. Relative to the MSE, and for errors of equal size, $S$ weights the squared error of an hour at 400~km~s$^{-1}$ by 0.67, at 600~km~s$^{-1}$ by 1.8 and in the top percentile of observed speeds by 18 ((Figure~\ref{fig:weights} in \ref{app:score})): it keeps every hour in the score while giving fast wind much more weight.

The training loss function includes boundary smoothness penalties,
\begin{equation}
 \mathcal{J}(\theta)=S\left(\mathcal{F}_{\mathrm{ACE}}(u_\theta),y\right)
 +\lambda_\phi\left\langle(\Delta_\phi^2u_\theta)^2\right\rangle
 +\lambda_t\left\langle(\Delta_t^2u_\theta)^2\right\rangle,
 \quad \lambda_\phi=\lambda_t=10^{-3}.
 \label{eq:loss}
\end{equation}
The penalties are added to $S$ at every optimization step. They are mean squared second differences of the boundary speed in longitude and time, in (km~s$^{-1}$)$^2$, while $S$ is in km~s$^{-1}$; with $\lambda=10^{-3}$ they act as a weak regularization rather than as a second objective of comparable weight. 

Because the weights of $\operatorname{TIMSE}_N$ grow with the observed speed, minimizing it does not lead to the average of the speeds that may follow given inputs, but to an average in which the faster of those speeds count more, and which therefore lies above it. The score $S$ combines this with the ordinary mean squared error, so its optimum lies between the two: at the optimum it acts as a squared error whose weight for each hour is proportional to $1/\mathrm{MSE}+w(y_i)/\operatorname{TIMSE}_N$. Training on $S$ therefore buys amplitude in fast wind at the cost of a positive bias, the familiar consequence of weighting errors by the outcome \cite{Gneiting2011,Lerch2017}. We accept this trade-off because the application gives priority to the reconstruction of fast streams at operational thresholds, and we quantify its cost with the MSE-trained variant (Section~\ref{sec:variants}) and with the bias-aware and threshold-free measures below.

Besides $S$ and the RMSE, we report the mean bias (predicted minus observed), the Pearson correlation $r$, and the amplitude ratio, the standard deviation of the predictions divided by that of the observations. For fast wind above a threshold $c$ of 500 or 600~km~s$^{-1}$ we use three measures. The true skill statistic, $\mathrm{TSS}=\mathrm{POD}-\mathrm{POFD}$, is the probability of detecting a fast hour minus the probability of a false detection, with predictions and observations both thresholded at $c$. The area under the ROC curve (AUC) uses the predicted speed as a score for the event $y\geq c$, and therefore does not depend on calibration. The best-threshold TSS is the TSS maximized over prediction cutoffs from 300 to 800~km~s$^{-1}$. Bias-removed RMSE and TSS are computed after subtracting each model's own mean bias from its predictions. Errors are also shown conditionally, in 25~km~s$^{-1}$ bins of the observed and of the predicted speed, and as the tail error $\sqrt{M_k}$ as a function of the threshold. Event-based verification of high-speed streams follows the protocol of Section~\ref{sec:cases}.

\subsection{Training, validation and reference methods}
\label{sec:train}
We evaluate all models by cross-validation over the five development batches (Section~\ref{sec:screening}). Each batch serves once as the test batch, scored only after training is complete; the batch preceding it in time serves as the validation batch (for the first batch, the last one), and the remaining three are used for training. Because the batches are contiguous in time, every test result comes from a period of the solar cycle that the model has not seen.

For training, each batch is cut into eight contiguous chunks of about 86 days. The solver needs a history of the boundary before it can produce speeds at ACE, so each chunk starts with a 150-h warm-up, initialized as described in \ref{app:numerics}, during which the solver runs but no hours are scored. Each optimization step uses the scored hours of eight chunks together. We use the Adam optimizer with an initial learning rate of $2\times10^{-3}$, reduced when progress stalls, and keep the model that scores best on the validation batch, stopping after 50 epochs without improvement or at most 300 epochs. No dropout or weight decay is used.

Each fold is trained twice, with different random initializations, giving ten models in total. The prediction for each test hour is the mean of the two models of its fold, so that every reported score is computed out of fold over all five test batches.

We compare the reference model with three simpler methods. The linear comparator is an ordinary least-squares regression of the speed on the eight features selected in Section~\ref{sec:screening}: the six AIA features of Table~\ref{tab:channels} and the two HMI features, the fraction of weak-field pixels ($|B_r|<10$~G) and the mean $|B_r|$, both within $7.5^\circ$ of the meridian and $30^\circ$ of the ecliptic. It uses the image taken four days before the target hour (Section~\ref{sec:screening}), and is fitted on the training batches only; unlike the reference model, it has access to magnetogram information. The 27-day recurrence predicts the speed observed one solar rotation ([27.27]~d) earlier, from the same ICME-masked series, and climatology predicts the mean speed of the training batches of the fold.

As a physics-based reference we use the Wang--Sheeley--Arge model (WSA) version 4.5 driven by GONG magnetograms \cite{Arge2000,Arge2003}, from the hourly archive of the Community Coordinated Modeling Center. At each hour of a test interval its boundary is the most recent WSA map, so that, like the reference model, it uses only information available at that time. The WSA speeds at ACE's latitude are decelerated to $21.5\,R_\odot$ as in HUXt and propagated to ACE with HUXt \cite{Owens2020} over the same intervals, with the same warm-up, and scored on the same hours. To tell whether differences come from the WSA boundary or from its propagation, we also propagate the same WSA boundary with our own solver, and map it ballistically to ACE. WSA is not tuned to these data: its parameters were set by its developers on earlier periods. 

To compare the methods on equal terms, all are scored on the same 46,412 matched hours: the test hours on which every method has a prediction. The other measured hours lack an input that one of the comparators needs, for example the image four days earlier or the speed one rotation earlier. Uncertainties of the differences between methods come from a block bootstrap of the matched hours, with blocks of 27 days to respect the correlation between consecutive hours (1000 resamples; the best-threshold TSS is re-optimized in each
resample). For the ablations, changes in $S$ are given as the mean and standard error over the ten trained models, with the number of folds in which the change has the same sign; the $p$-values test the five fold means, each averaged over the two seeds, with a paired $t$-test.

All results are reconstructions: they use the images and other inputs available when the relevant plasma leaves the inner boundary, and are not forecasts issued at a fixed time. Comparisons with published forecasts issued at fixed times, on held-out data, are the subject of Part~II.

\subsection{Model variants}
\label{sec:variants}
The reference model combines three choices: the training loss $S$, the transport solver as the link between the boundary and ACE, and a neural network as the boundary map. To find out what each contributes, we train variants that change one of these choices at a time and keep everything else as in the reference model: the inputs, the five folds, the two seeds and the training protocol (Section~\ref{sec:train}). Table~\ref{tab:variants} lists them.

The first variant changes only the loss, replacing $S$ by the mean squared error; it shows what the speed-weighted score buys and what it costs. Two variants change only the link to ACE, using the simpler forward operators defined in Section~\ref{sec:corotating}: ballistic propagation, in which every parcel travels at its own speed without interacting with its neighbours, and a fixed 96-h lag, in which the travel time does not depend on the speed at all. Comparing them with the solver separates the effect of a speed-dependent travel time from that of stream interaction.

The remaining variants change the boundary map, to test how much model flexibility is needed. The simplest is a linear map: a weighted sum of the 37 inputs plus a constant, passed through the same output sigmoid as the network (38 weights). We train it through the solver and, for comparison, through the fixed lag. Between the linear map and the network lies an additive model, a neural additive model (NAM; \citeA{Agarwal2021}). In a linear map, each input changes the boundary speed in proportion to its value; in an additive model, each input contributes through its own learned curve, which may bend, and the contributions of all inputs are added. The effect of every input can therefore still be read off separately, as a curve, which a general neural network does not allow. Our NAM keeps a linear term in all 37 inputs and adds a learned curve for each of the eight current-time inputs (the six image features, $f_s$ and the recurrence). It also adds a learned surface for each of three pairs of inputs whose combined effect may matter, so that the effect of one can depend on the value of the other: coronal-hole area with the recurrence, coronal-hole area with $f_s$, and the hole-boundary distance with the recurrence.

Both the linear map and the NAM are readable in principle, but not with these inputs: they are strongly correlated, since the lagged copies of a feature, and the three area features, carry nearly the same information, so the data constrain only the combined effect of each group, and different runs divide it differently among the inputs. Two further variants therefore reduce the inputs to one value per physical quantity, the mean of the current and the available lagged values, and keep four quantities: coronal-hole area (channel~0), the dark fraction near the ecliptic (channel~1), the darkness of the dark pixels (channel~3) and the back-mapped recurrence, with the availability flags. The four-coefficient model is the linear map of these quantities, with a single coefficient each (10 weights); the four-curve model is the corresponding additive model, with a smooth learned curve for each quantity and a penalty on its curvature [$\lambda=0.3$] (911 weights).

\begin{table}[t]
\caption{Trained models. All use the inputs of Table~\ref{tab:channels} (or the reduced set where stated), the five folds and two seeds, and the protocol of Section~\ref{sec:train}.}
\label{tab:variants}\centering\small
\begin{tabular}{llllr}\hline
Model & Boundary map & Link to ACE & Loss & Parameters\\\hline
Reference model & MLP & solver & $S$ & 12,377\\
Reference model, MSE loss & MLP & solver & MSE & 12,377\\
Ballistic link & MLP & ballistic, no interaction & $S$ & 12,377\\
Fixed lag & MLP & fixed 96-h lag & $S$ & 12,377\\
Linear map & linear & solver & $S$ & 38\\
Linear map, fixed lag & linear & fixed 96-h lag & $S$ & 38\\
NAM & linear + 8 curves + 3 surfaces & solver & $S$ & 6,833\\
Four-coefficient model & linear, 4 quantities & solver & $S$ & 10\\
Four-curve model & 4 curves & solver & $S$ & 911\\\hline
\end{tabular}
\end{table}

\section{Results}
\label{sec:results}
Unless stated otherwise, all scores below are cross-validated development estimates pooled over the 46,412 matched test hours, and the prediction of each trained model is the mean of its two seeds. Differences quoted as significant have 95\% paired block-bootstrap intervals that exclude zero (Section~\ref{sec:train}); the intervals are given in the tables.

\subsection{Reconstruction accuracy and amplitude}
\label{sec:accuracy}
Table~\ref{tab:performance} compares the reference model with the other methods on the same hours. On the combined score $S$, it clearly outperforms the linear comparator, 27-day recurrence and climatology, and each of its two training runs alone gives nearly the same result. Part of its advantage over the linear comparator comes from the near-zero bias of that regression, which $S$ penalizes (Section~\ref{sec:metric}); but even when the linear comparator's predictions are shifted by the constant that minimizes its $S$, the reference model remains about 10~km~s$^{-1}$ better.

\begin{figure*}[t]
\centering\includegraphics[width=\textwidth]{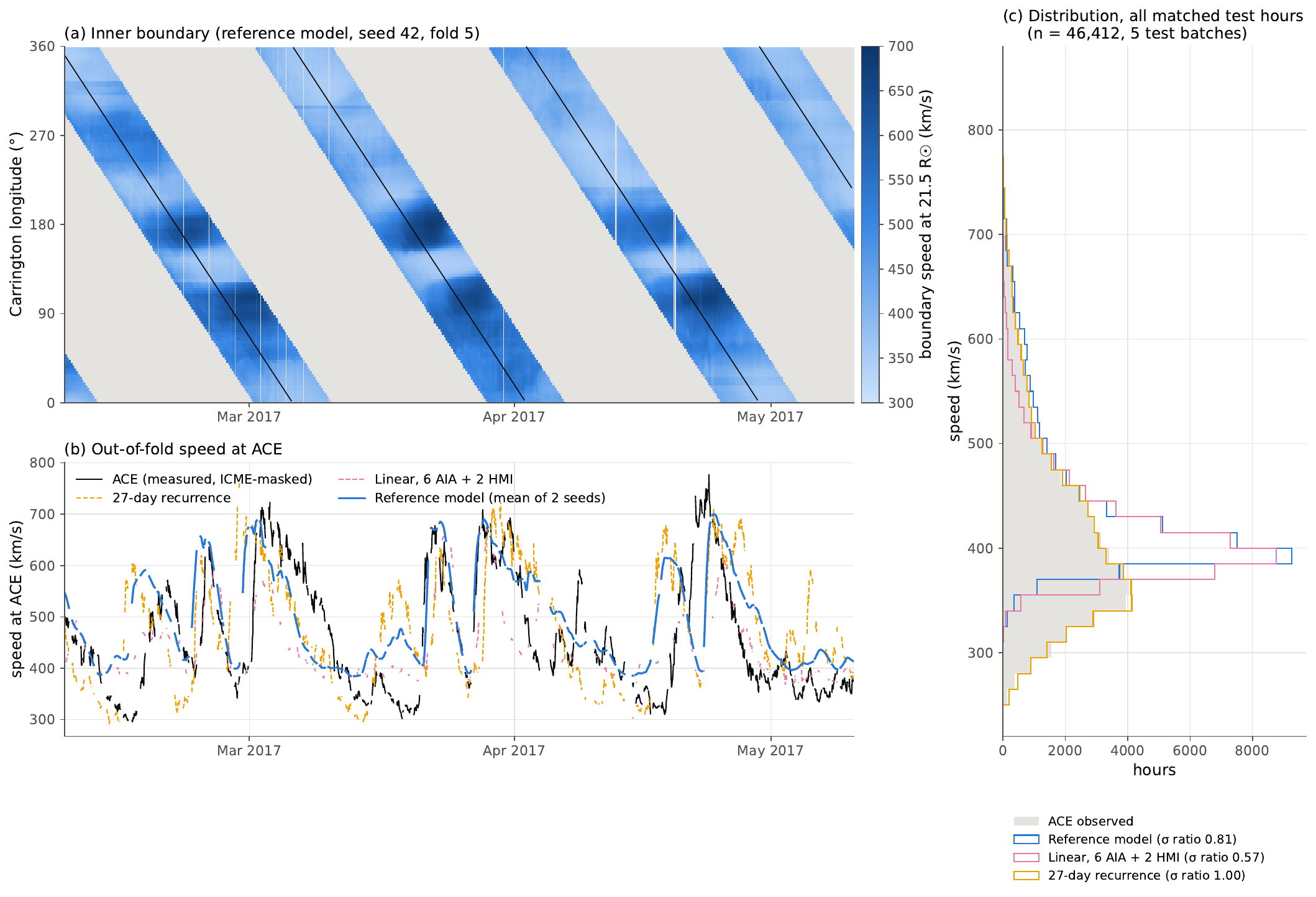}
\caption{Out-of-fold reconstruction. (a) Inferred inner-boundary speed for one test interval (2017 February 7 to May 10, declining phase; the reference model, seed 42, fold 5) per Carrington longitude and hour; grey: no AIA input; black: sub-Earth Carrington longitude. Only the cells along the sub-Earth track are constrained by ACE; elsewhere the boundary is the network's prediction from the same features. (b) Speed at ACE over the same interval: measured (ICME-masked), the reference model (mean of the two seeds), the linear comparator and 27-day recurrence; gaps are unscored hours. (c) Distribution of the speeds over all 46,412 matched test hours, with the ratio of predicted to observed standard deviation.}
\label{fig:reconstruction}
\end{figure*}

The reference model does not improve every measure. Its RMSE is higher than that of the linear comparator, while its correlation with the observations is higher and its predictions reproduce much more of the observed variability. The difference in RMSE is due to its positive bias of 26~km~s$^{-1}$: once each model's mean bias is removed, the two RMSEs are nearly equal (Table~\ref{tab:skill}). Conditioned on the observed speed, the errors of every model show regression toward the mean, with the slowest wind over-predicted and the fastest under-predicted (Figure~\ref{fig:condrmse}b); the mean bias is the offset that the training score adds to this pattern (Section~\ref{sec:hss}). Figure~\ref{fig:reconstruction} shows an example interval and the distribution of the predicted speeds.

The physics-based reference, WSA propagated with HUXt, scores much worse on the same hours: level with 27-day recurrence and behind the linear comparator, and behind the reference model in each of the five test batches. The difference lies in the WSA boundary rather than in its propagation, since the same boundary propagated with our solver gives nearly the same score. WSA over-predicts fast wind: it predicts speeds of at least 600~km~s$^{-1}$ in about twice as many hours as were observed, with large errors in those hours, and it discriminates fast wind less well than the reference model (AUC at 600~km~s$^{-1}$ 0.78 against 0.88). Recalibrating WSA linearly on the training batches removes its bias but compresses its predictions, so that it then hardly detects fast wind at fixed thresholds (Table~\ref{tab:skill}).

\begin{table}[t]
\caption{Cross-validated performance on the 46,412 matched development test hours (trained models: mean of two seeds). $S$, RMSE and bias in km~s$^{-1}$; $\Delta S$: difference from the reference model, with 95\% paired block-bootstrap interval; Amp.: ratio of predicted to observed standard deviation; TSS and AUC for fast wind at 500 / 600~km~s$^{-1}$. The linear comparator uses six AIA and two HMI features (Section~\ref{sec:train}). Bold: best value in each column.}
\label{tab:performance}\centering\footnotesize
\begin{tabular}{lrcrrrrcc}\hline
Method & $S$ & $\Delta S$ [95\%] & RMSE & Bias & $r$ & Amp. & TSS$_{500/600}$ & AUC$_{500/600}$\\\hline
Reference model ($S$ loss) & \textbf{93.7} & --- & 79.3 & $+26.4$ & 0.58 & 0.81 & \textbf{0.46} / \textbf{0.38} & \textbf{0.816} / \textbf{0.879}\\
Reference model, MSE loss & 99.5 & $+5.7$ [3.4, 8.1]  & \textbf{71.6} & $+0.1$ & \textbf{0.59} & 0.63 & 0.36 / 0.15 & 0.809 / 0.870\\
Linear comparator & 107.0 & $+13.2$ [7.9, 17.4] & 75.8 & $+3.1$ & 0.52 & 0.57 & 0.28 / 0.10 & 0.799 / 0.850\\
Linear comparator, shifted & 103.9 & $+10.2$ [4.8, 14.4] & --- & --- & 0.52 & 0.57 & --- & 0.799 / 0.850\\
27-day recurrence & 119.6 & $+25.9$ [19.5, 31.8] & 91.9 & $-0.7$ & 0.46 & \textbf{1.00} & 0.34 / 0.26 & 0.760 / 0.802\\
Climatology & 137.5 & $+43.7$ [37.3, 50.5] & 89.5 & \textbf{0.0} & --- & --- & 0 / 0 & ---\\\hline
WSA 4.5 + HUXt & 120.3 & $+26.6$ [17.2, 35.9] & 112.3 & $+44.0$ & 0.34 & 1.03 & 0.26 / 0.29 & 0.688 / 0.778\\
WSA 4.5 + ballistic & 126.3 & $+32.6$ [23.7, 41.1] & 133.5 & $+79.0$ & 0.28 & 1.04 & 0.24 / 0.33 & 0.666 / 0.764\\\hline
\end{tabular}
\end{table}

% ----------------------------------------------------------------------
\subsection{Fast wind and the training objective}
\label{sec:hss}
The two training objectives reward different things, and comparing the reference model with the same network trained on the mean squared error shows what this means in practice (Table~\ref{tab:performance}). Trained on the mean squared error, the network gives the most accurate speeds on average of all methods, without bias, and is clearly more accurate than the linear comparator. Its predictions, however, are compressed: he spread of their predictions is only 63\% of the observed spread, and at a threshold of 600~km~s$^{-1}$ they detect few of the fast-wind hours. Trained on $S$, the same network its predictions spread to 81\% of the observed spread and detects fast wind far more often, at the price of a positive bias and a higher RMSE.

Because both $S$ and the TSS at a fixed threshold reward over-forecasting (Section~\ref{sec:metric}; \citeA{Gneiting2011,Lerch2017}), we also compare the models with measures that do not depend on calibration (Table~\ref{tab:skill}). They tell a consistent story. The ability to rank fast wind above slow wind, measured by the AUC, is nearly the same for both losses, and slightly but significantly better than that of the linear comparator, as is the best TSS over all decision thresholds. The large difference at the fixed threshold of 600~km~s$^{-1}$, a TSS of 0.38 against 0.10, is therefore mostly a difference in amplitude: the linear comparator rarely predicts speeds that high at all, and after bias removal the gap narrows considerably. The loss changes how far the predictions reach, not how well they rank.

\begin{figure*}[t]
\centering\includegraphics[width=\textwidth]{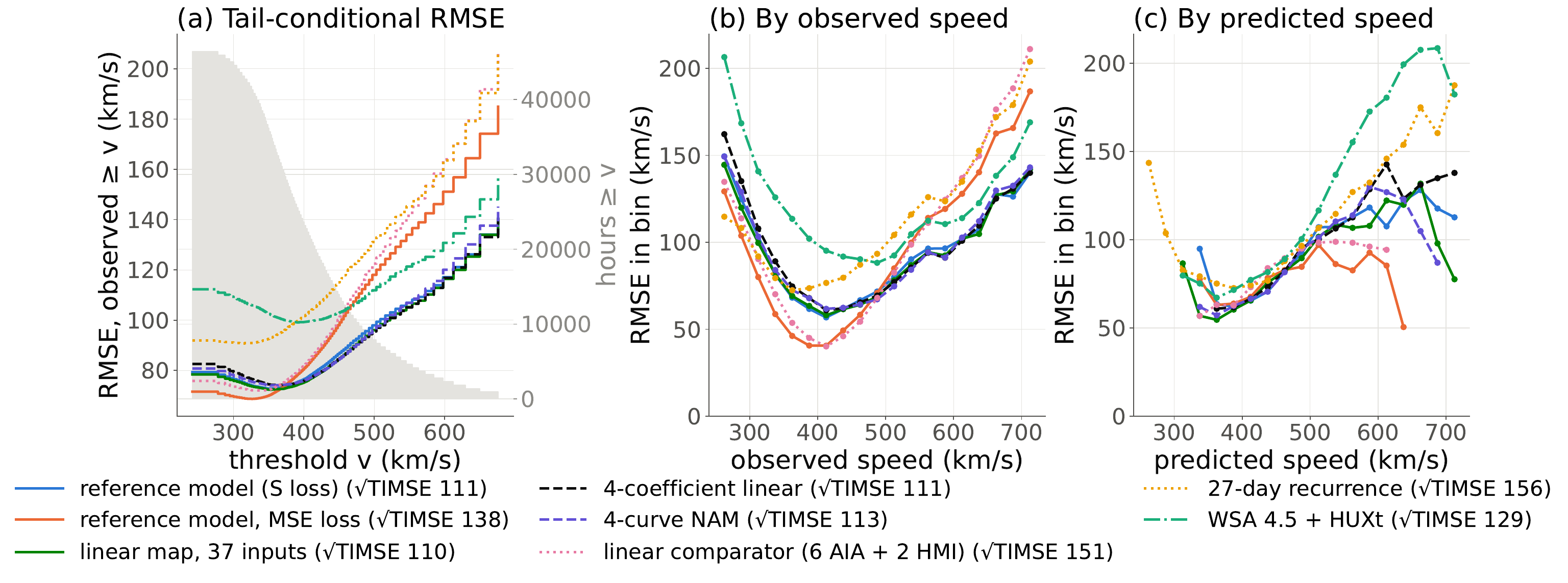}
\caption{Conditional errors on the matched development test hours (out of fold, mean of two seeds). (a) RMSE over the hours whose observed speed is at least $v$, as a function of $v$; the area under the squared curve, divided by the speed range, is $\operatorname{TIMSE}_N$, the tail term of $S$. Grey: number of hours. (b) RMSE in 25~km~s$^{-1}$ bins of the observed speed. (c) RMSE in bins of each model's own predicted speed (bins with at least 100 hours).}
\label{fig:condrmse}
\end{figure*}

The conditional errors show where the two objectives differ (Figure~\ref{fig:condrmse}). The models trained on $S$ make larger errors in slow wind, which they over-predict, and much smaller errors in fast wind; their tail-conditional curves cross those of the MSE-trained model near 400~km~s$^{-1}$. When an $S$-trained model predicts fast wind, its prediction is individually less certain, but the MSE-trained model rarely makes such predictions in the first place.

We adopt the model trained on $S$ because the application is the reconstruction of fast streams at operational thresholds. Users who need unbiased speeds can subtract its mean bias, which brings its RMSE close to that of the linear comparator, or use the MSE-trained variant.

\begin{table}[t]
\caption{Bias-aware and threshold-free measures on the matched hours, at 500 / 600~km~s$^{-1}$. Bias removed: each method's own mean bias subtracted from its predictions. Best TSS: maximized over prediction cutoffs from 300 to 800~km~s$^{-1}$. AUC: area under the ROC curve, which does not depend on calibration. Recalibrated WSA: $a+b\,v_{\mathrm{WSA}}$ fitted on each fold's training batches. Bottom rows: differences with 95\% paired block-bootstrap intervals. Bold: best value in each column.}
\label{tab:skill}\centering\footnotesize
\begin{tabular}{lcccc}\hline
Method & RMSE, bias removed & TSS, bias removed & Best TSS & AUC\\\hline
Reference model ($S$ loss) & 74.8 & \textbf{0.42} / \textbf{0.29} & \textbf{0.51} / \textbf{0.63} & \textbf{0.816} / \textbf{0.879}\\
Reference model, MSE loss & \textbf{71.6} & 0.36 / 0.15 & 0.49 / 0.62 & 0.809 / 0.870\\
Linear comparator & 75.8 & 0.27 / 0.09 & 0.47 / 0.56 & 0.799 / 0.850\\
27-day recurrence & 91.9 & 0.34 / 0.26 & 0.42 / 0.48 & 0.760 / 0.802\\
WSA 4.5 + HUXt & 103.3 & 0.23 / 0.19 & 0.27 / 0.42 & 0.688 / 0.778\\
WSA 4.5 + HUXt, recalibrated & 84.1 & 0.03 / 0.00 & 0.25 / 0.39 & 0.675 / 0.759\\\hline
\multicolumn{5}{l}{\emph{Differences}}\\
Reference $-$ linear comparator & --- & --- & \parbox[t]{2.6cm}{\centering $+0.044$ [0.012, 0.076] /\\ $+0.067$ [0.023, 0.116]} & \parbox[t]{2.6cm}{\centering $+0.017$ [0.003, 0.031] /\\ $+0.029$ [0.008, 0.050]}\\
Reference $-$ MSE loss & --- & --- & --- &$+0.007$ [$-0.001$, 0.017] / $+0.009$ [$-0.003$, 0.023]\\
Reference $-$ WSA + HUXt & --- & --- & --- & $+0.128$ [0.100, 0.157] / $+0.101$ [0.062, 0.141]\\\hline
\end{tabular}
\end{table}

\begin{figure*}[t]
% TODO (figure script): keep only panel (a); panels (b) and (c) repeat the TSS and AUC values of Table~\ref{tab:performance}. Then shorten the caption.
\centering\includegraphics[width=0.62\textwidth]{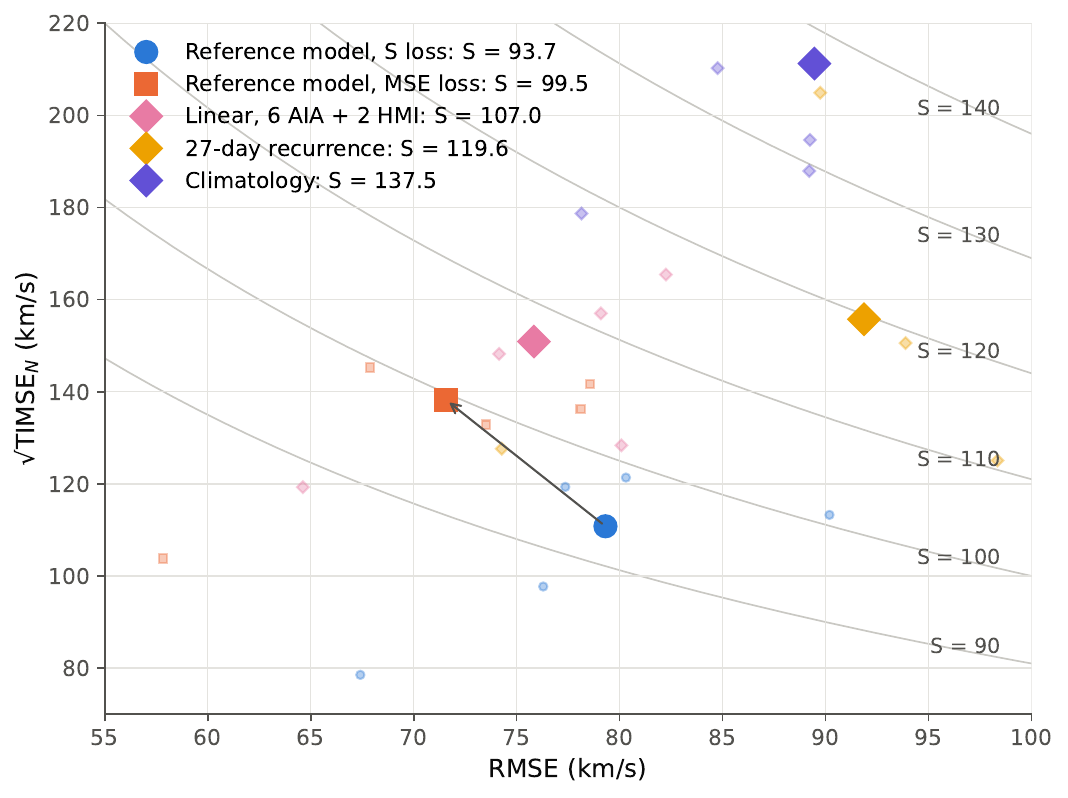}
\caption{What the loss buys, on the 46,412 matched development test hours (trained models: mean of the two seeds): RMSE
against the tail error $\sqrt{\operatorname{TIMSE}_N}$, pooled over the five test batches (large markers) and per batch
(small markers); grey curves are lines of constant $S$. The arrow joins the reference model trained with $S$ to the same
model trained with the mean squared error.}
\label{fig:hss}
\end{figure*}

% ----------------------------------------------------------------------
\subsection{What the link to the spacecraft adds}
\label{sec:solveradds}
The reference model differs from the linear comparator in four ways at once: its inputs, its training score, its model class and the link from the boundary to the spacecraft. To find out which of these matters, we changed one at a time (Table~\ref{tab:factorial}).

The link matters most. Trained through a fixed 96-h lag instead of the solver, the network loses 8.6~km~s$^{-1}$ in $S$ and much of its ability to discriminate fast wind, and the linear map loses about as much. This is the largest single part of the reference model's advantage over the linear comparator. Most of the remainder comes from the inputs and the training score, which bring the linear comparator's model class, still at a fixed lag, about halfway to the reference model; the nonlinearity of the network adds nothing measurable (Section~\ref{sec:modelclass}).

\begin{figure*}[t]
\centering\includegraphics[width=\textwidth]{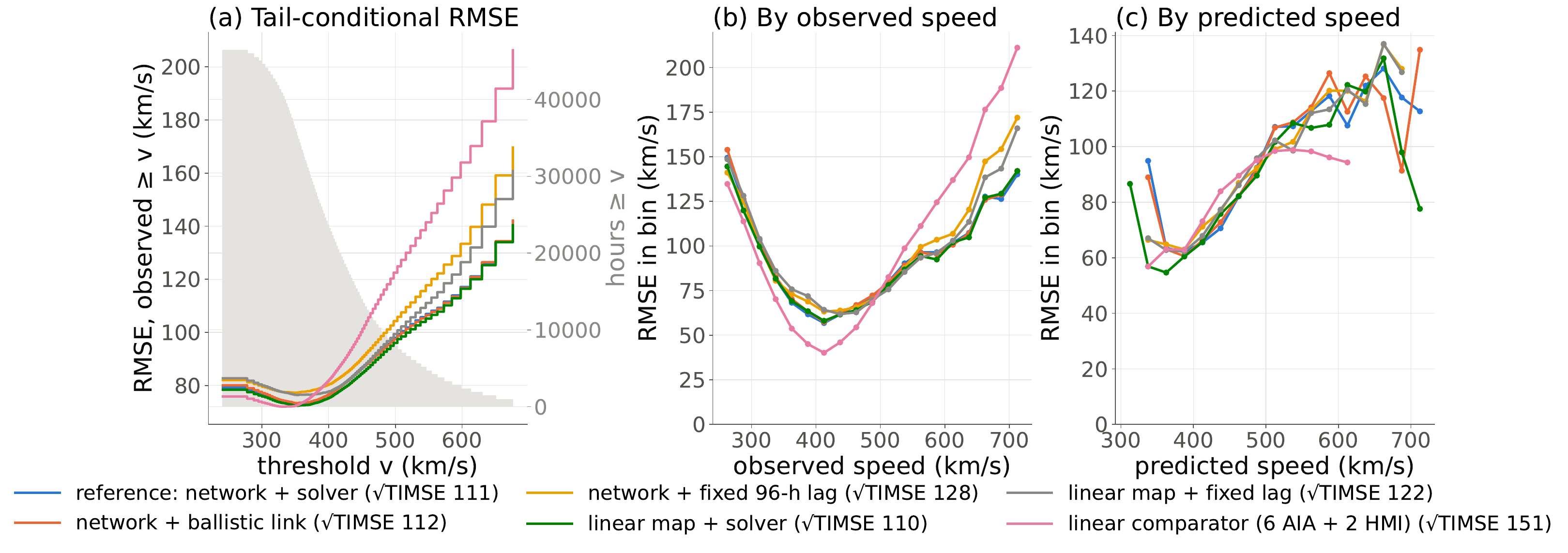}
\caption{Conditional errors of the link variants, as in Figure~\ref{fig:condrmse}: the reference model (network and solver), the network trained through the ballistic link and through the fixed 96-h lag, the linear map trained through the solver and through the fixed lag, and the linear comparator. All variants are trained end to end through their link.}
\label{fig:condrmse_links}
\end{figure*}

The fixed lag does not lose at the steep stream interfaces, where both links give the same error, but in the rest of the series: in slow wind of 350--425~km~s$^{-1}$, and above all in fast wind above 625~km~s$^{-1}$, which the fixed-lag model under-predicts more strongly (Figure~\ref{fig:condrmse_links}). The speed-dependent travel time thus matters most for the levels of slow and fast streams, rather than for the transitions between them.

Whether that travel time comes from the full transport solver or from simple ballistic propagation makes little difference. Trained through a ballistic link, in which each parcel travels at its own speed, the network reaches the same score and the same discrimination as with the solver; the two are equal in fast wind and at the stream interfaces, and the solver is slightly better only in slow wind. In the event-based verification (Section~\ref{sec:cases}) the trained variants do not differ in threat score, although those trained at a fixed lag produce more false alarms and larger timing errors.

\begin{table}[t]
\caption{Trained variants (Table~\ref{tab:variants}) on the 46,412 matched hours (mean of two seeds). $\Delta S$: difference from the reference model with 95\% paired block-bootstrap interval; $S$, RMSE and bias in km~s$^{-1}$; Amp.: amplitude ratio; AUC and TSS at 500 / 600~km~s$^{-1}$. Bold: best value in each column.}
\label{tab:factorial}\centering\small
\begin{tabular}{lrrrrccc}\hline
Model & $S$ & $\Delta S$ [95\%] & RMSE & Bias & Amp. & AUC & TSS\\\hline
\multicolumn{8}{l}{\emph{Network, different links}}\\
Reference model (solver) & 93.7 & --- & 79.3 & $+26.4$ & 0.81 & 0.816 / 0.879 & 0.46 / 0.38\\
Ballistic link & 94.5 & $+0.7$ [$-0.4$, 2.0] & 80.1 & $+28.1$ & 0.82 & 0.817 / 0.881 & 0.45 / \textbf{0.41}\\
Fixed 96-h lag & 102.4 & $+8.6$ [3.9, 12.4] & 82.0 & $+26.2$ & 0.79 & 0.791 / 0.838 & 0.42 / 0.27\\
\multicolumn{8}{l}{\emph{Other boundary maps}}\\
Linear map, solver & \textbf{92.9} & {\boldmath$-0.8$} [$-2.4$, 0.8] & 78.4 & $+24.4$ & 0.83 & \textbf{0.826} / \textbf{0.886} & 0.46 / 0.36\\
Linear map, fixed lag & 100.3 & $+6.6$ [1.9, 10.8] & 82.7 & $+31.3$ & 0.80 & 0.805 / 0.851 & 0.45 / 0.32\\
NAM, solver & 94.1 & $+0.3$ [$-1.8$, 2.5] & 80.9 & $+25.5$ & \textbf{0.85} & 0.815 / 0.884 & 0.46 / 0.35\\
Four-curve model & 95.6 & $+1.8$ [$-0.4$, 3.9] & 80.8 & $+27.7$ & 0.81 & 0.817 / 0.870 & 0.46 / 0.33\\
Four-coefficient model & 95.6 & $+1.9$ [$-0.9$, 4.7] & 82.6 & $+33.0$ & 0.80 & 0.817 / 0.874 & \textbf{0.47} / 0.35\\
\multicolumn{8}{l}{\emph{Reference}}\\
Linear comparator & 107.0 & $+13.2$ [7.9, 17.4] & \textbf{75.8} & {\boldmath$+3.1$} & 0.57 & 0.799 / 0.850 & 0.28 / 0.10\\\hline
\end{tabular}
\end{table}

% ----------------------------------------------------------------------
\subsection{Model class, inputs and interpretability}
\label{sec:modelclass}
\label{sec:ablations}
A linear function of the reference model's 37 inputs, trained through the solver in the same way, scores as well as the network and discriminates fast wind slightly better (Table~\ref{tab:factorial}). The more flexible models fit the training batches better but gain nothing on validation or test data. The boundaries produced by the network and by the linear map differ by 23~km~s$^{-1}$ RMS, no more than the boundaries of two training runs of the same model (16--23~km~s$^{-1}$): ACE constrains the boundary only to this tolerance, and within it a linear function of the inputs suffices. The additive model (NAM), with a learned curve for each key input and three pair interactions, also scores as well as the network.

The ablations of Table~\ref{tab:ablations} show which inputs the model relies on. Removing the back-mapped recurrence is the only change that clearly degrades it. Adding the two HMI features does not help, and removing the expansion factor $f_s$ changes the score by less than the scatter between training runs; we keep $f_s$ for its physical role in the WSA relation. Daily instead of hourly expansion factors, a smaller network, dropout and weight decay make no difference: the reference model is not tuned to a sharp optimum.

Since a simple map does as well as the network, the learned boundary can be expressed in physical terms. The 37 inputs themselves are too correlated for that, so we reduced them to one value per physical quantity and kept four quantities: coronal-hole area, the dark fraction near the ecliptic, the darkness of the dark pixels and the back-mapped recurrence (Section~\ref{sec:variants}). We fitted two versions, which differ in how each quantity acts on the boundary speed. In the four-coefficient model, each quantity changes the speed in proportion to its value, through a single coefficient. In the four-curve model, an additive model, each quantity acts through its own smooth learned curve, so its effect may bend or level off. Both come within the uncertainty of the network, slightly behind the linear map with all inputs (Table~\ref{tab:factorial}).

The four coefficients have the same sign in all ten models trained on different folds and seeds, except for the weakest, and their sizes are physically plausible (Table~\ref{tab:coefficients}). Coronal-hole area dominates: one standard deviation of area raises the boundary speed by about 32~km~s$^{-1}$, against about 19~km~s$^{-1}$ for the recurrence and 16~km~s$^{-1}$ for the darkness of the coronal-hole pixels; once area is included, the dark fraction near the ecliptic adds little.

\begin{figure*}[t]
\centering\includegraphics[width=\textwidth]{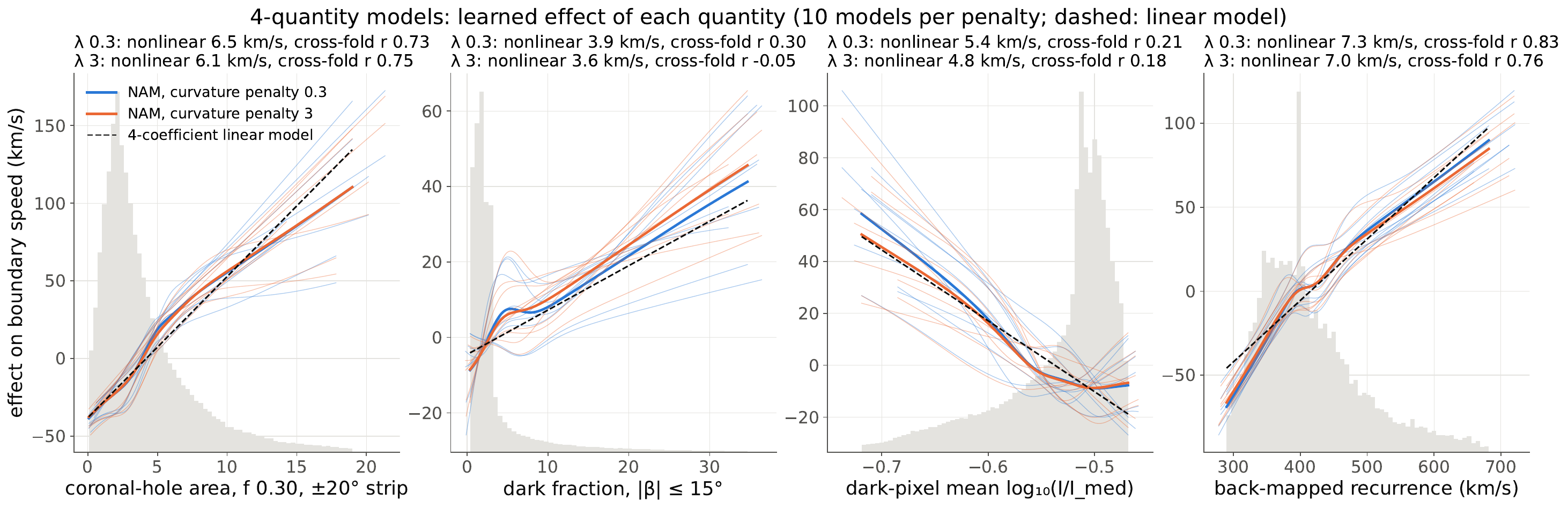}
\caption{Learned effect of each of the four physical quantities on the boundary speed (at 400~km~s$^{-1}$), for the four-curve model with two strengths of the curvature penalty (thin lines: the ten models of each; thick: their mean) and for the four-coefficient model (dashed). Grey: distribution of each quantity over the boundary cells. Panel headers give the nonlinear part of the curves (RMS deviation from their best straight line) and its correlation between models trained on different folds.}
\label{fig:qnam}
\end{figure*}

The curves of the four-curve model (Figure~\ref{fig:qnam}) show how far the linear picture holds. Only one curve has a shape that recurs in models trained on different folds: the boundary speed rises steeply with coronal-hole area at small and moderate areas and levels off for the largest holes. The recurrence curve also recurs, but its main feature, a step at 400~km~s$^{-1}$, is an artefact of the value used where the recurrence is missing; the curves of the other two quantities do not recur. A piecewise-linear model whose bends were free to move kept them near where they started, and scored the same from every start: the data allow the curves to bend, but do not say where.

The levelling-off with coronal-hole area is nevertheless real. For the largest holes, in the top 2.5\% of hours, the four-coefficient model over-predicts by about 100~km~s$^{-1}$, and the models that level off, the reference model and the four-curve model, are about 20~km~s$^{-1}$ more accurate there. The global score does not show this, because lowering the predictions for large holes removes over-predictions, which lowers the RMSE, but also lowers the predictions for the large-hole hours that were fast, which the tail term of $S$ penalizes; the two effects cancel.

\begin{table}[t]
\caption{One-change ablations of the reference model. $\Delta S$ = $S$(variant) $-$ $S$(reference model) in km~s$^{-1}$, paired by fold and seed; mean $\pm$ standard error over ten runs; $p$ from a fold-level paired $t$-test (seeds averaged per fold, 4 degrees of freedom). Bold: lowest $\Delta S$ in each column.}
\label{tab:ablations}\centering\small
\begin{tabular}{lcc}\hline
Change to the reference model & $\Delta S$, test & $\Delta S$, validation\\\hline
No back-mapped recurrence & $+4.8\pm1.3$ ($p=0.05$) & $+4.2\pm0.7$ ($p=0.01$)\\
Add the two HMI features & $+1.9\pm1.2$ ($p=0.27$) & $+0.4\pm0.6$ ($p=0.65$)\\
No expansion factor $f_s$ & $+0.3\pm0.9$ ($p=0.84$) & $+1.1\pm0.5$ ($p=0.18$)\\
Daily instead of hourly $f_s$ & $+0.2\pm0.2$ & $+0.0\pm0.2$\\
Smaller network (64--32--16) & $+0.2\pm0.4$ & $+0.4\pm0.3$\\
\end{tabular}
\end{table}

\begin{table}[t]
\caption{Coefficients of the four-coefficient model: change of the boundary speed (evaluated at 400~km~s$^{-1}$) per standard deviation of each quantity (mean $\pm$ standard deviation over the ten models of five folds $\times$ two seeds) and per physical unit, and the number of models sharing the sign shown. Area and dark fraction in per cent of the pixels in their strip; darkness: 0.1~dex lower intensity of the dark pixels.}
\label{tab:coefficients}\centering\footnotesize
\begin{tabular}{lccc}\hline
Quantity & Per s.d. & Per unit & Sign\\
 & (km~s$^{-1}$) & (km~s$^{-1}$) & \\\hline
Coronal-hole area & $+32\pm2$ & $+9.5$ per \% & 10/10\\
Back-mapped recurrence & $+19\pm3$ & $+0.35$ per km~s$^{-1}$ & 10/10\\
Darkness of dark pixels & $+16\pm7$ & $+27$ per 0.1~dex & 10/10\\
Ecliptic dark fraction & $+8\pm6$ & $+1.2$ per \% & 9/10\\\hline
\end{tabular}
\end{table}

\subsection{High-speed-stream events and case studies}
\label{sec:cases}
To examine individual streams, we applied the event-based verification protocol of \citeA{Collin2026}, which follows \citeA{Owens2005} and \citeA{Reiss2016}, to the out-of-fold predictions. The observed high-speed-stream peaks are taken from their published enhancement catalogue, derived from OMNI data, and the same peak detector is applied to each prediction: the series is smoothed over 24~h, and peaks above 390~km~s$^{-1}$ with a prominence of at least 35~km~s$^{-1}$, separated by at least four days, are retained. A predicted peak within three days of an observed peak counts as a hit; unmatched observed peaks are misses, and unmatched predicted peaks are false alarms. The comparison uses the 51,928 OMNI hours on which the reference model, the linear comparator and recurrence are all available.

Of 187 observed streams, the reference model detects 141, with 24 false alarms; the linear comparator detects 125, and recurrence 128 with twice as many false alarms (Table~\ref{tab:events}). The largest difference is in the speed of the detected peaks, which the reference model reproduces far better than the linear comparator; this reflects amplitude, that is the inputs and the training score, since all trained variants of Table~\ref{tab:factorial} do about equally well. For the detected streams, the median timing error is 20~h and the median error in peak speed 44~km~s$^{-1}$.

\begin{table}[t]
\caption{High-speed-stream event verification on the development test batches (protocol of \protect\citeA{Collin2026}; 187 observed streams; 51,928 OMNI hours). POD: probability of detection; TS: threat score; peak RMSE: RMSE of the speed at matched peaks. Bold: best value in each column.}
\label{tab:events}\centering
\begin{tabular}{lrrrrrr}\hline
Method & Hits & Misses & False alarms & POD & TS & Peak RMSE (km~s$^{-1}$)\\\hline
Reference model & \textbf{141} & \textbf{46} & 24 & \textbf{0.75} & \textbf{0.67} & \textbf{74}\\
Linear comparator & 125 & 62 & \textbf{23} & 0.67 & 0.60 & 123\\
27-day recurrence & 128 & 59 & 47 & 0.68 & 0.55 & 86\\\hline
\end{tabular}
\end{table}

\begin{figure*}[t]
\centering\includegraphics[width=\textwidth]{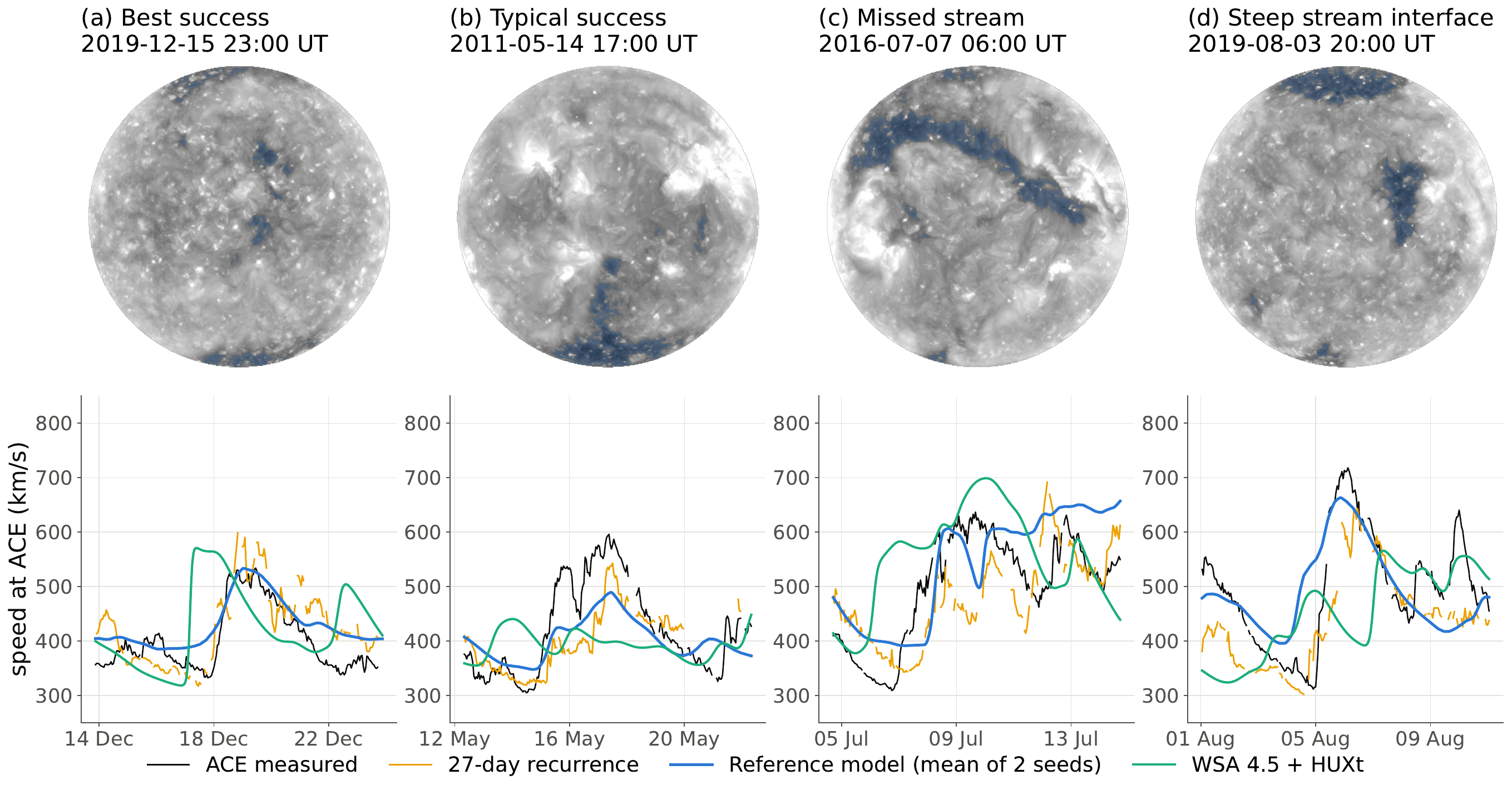}
\caption{Case studies selected by rule from the out-of-fold high-speed-stream events: (a) best success, (b) typical success (median error), (c) largest missed stream, and (d) steepest stream interface, with observed peaks on 2019 December 18, 2011 May 17, 2016 July 9 and 2019 August 6. Top: AIA 193~\AA{} frame at the departure time of the observed peak from $21.5\,R_\odot$, dark pixels in blue. Middle: inferred inner-boundary speed (the reference model, seed 42), with the sub-Earth Carrington longitude (black) and the frame time (dashed). Only the cells along the sub-Earth track are constrained by ACE; elsewhere the boundary is the network's prediction from the same features. Bottom: measured ACE speed, the reference model (mean of two seeds), the linear comparator and 27-day recurrence; triangles mark the observed and predicted peaks. In (c) the predicted series has no distinct peak, so the event counts as a miss although the onset is reproduced. In (d) the solver (seed 42) is compared with a ballistic mapping of the same boundary; grey dots mark hours where ballistic parcels overtake one another and the mapping has several values.}
\label{fig:cases}
\end{figure*}

Four cases were selected from the event list of the reference model by rules fixed in advance (Figure~\ref{fig:cases}). For each detected stream, the timing and peak-speed errors were divided by their medians and added. The best success is the detection with the smallest combined error; the typical success, the detection with the median combined error; the failure, the missed stream with the highest observed peak; and the interface case, the detection with the steepest observed rise in speed (largest 24-h increase in the three days before the peak). We required at least 70\% data coverage within five days of each peak, and windows that do not overlap.

In the best success (peak on 2019 December 18 at 21~UT, 540~km~s$^{-1}$), the AIA image of 15 December shows small, isolated dark regions at low latitude near the central meridian; the reference model predicts the peak 5~h late and 7~km~s$^{-1}$ low, while the linear comparator stays near 420~km~s$^{-1}$. In the typical success (2011 May 17 at 08~UT, 598~km~s$^{-1}$), a coronal hole extends from the south polar region toward the equator near the meridian; the reference model times the stream well, 3~h late, but underestimates its peak by 109~km~s$^{-1}$, the usual failure being in amplitude rather than timing. In the largest missed stream (2016 July 9 at 17~UT, 638~km~s$^{-1}$), a large, elongated hole crosses the northern hemisphere and reaches the equator. The reference model predicts the rise on 8 July but keeps rising to about 650~km~s$^{-1}$ for five more days, so that its series has no distinct peak within three days and the event counts as a miss: the extended dark area is read as one long fast stream, whereas the observed stream decays after two days. In the steepest interface (2019 August 6 at 01~UT, 722~km~s$^{-1}$, after a rise of 398~km~s$^{-1}$ in 24~h), the source is a compact equatorial hole at the meridian, and the reference model predicts the peak 5~h early and 59~km~s$^{-1}$ low.

\subsection{Bias–variance–noise decomposition}
\label{sec:budget}
The noise estimate of Section~\ref{sec:degeneracy} allows the error of each model to be divided into three parts: the noise $\mathrm{E}[\mathrm{Var}(Y\mid X)]$ that no function of the inputs can remove; the variance due to training, estimated from the difference between the two seeds; and the bias, the remainder, which we split further into the mean offset and a ``shape'' part that varies with the conditions (Table~\ref{tab:budget}). Since the noise is estimated on all 65,047 development hours with inputs, the RMSE values here are out-of-fold values over those hours and differ slightly from Table~\ref{tab:performance}.

\begin{table}[t]
\caption{Bias–variance–noise decomposition over the 65,047 development hours with inputs (out of fold, mean of two seeds), in km~s$^{-1}$ RMS. Noise from the Gamma test (Section~\ref{sec:degeneracy}); variance from the seed differences (a lower bound); bias as the remainder, split into the mean offset and the shape part. Shares are of the mean squared error. The four-quantity models use less information than the 37 inputs, so their bias includes what they leave out. Bold: lowest value in each column (offset: closest to zero).}
\label{tab:budget}\centering\small
\begin{tabular}{lrrrrrrc}\hline
Model & RMSE & Noise & Variance & Bias & Offset & Shape & Noise / var. / bias\\\hline
Reference model ($S$ loss) & 79.8 & 67.6 & 8.0 & 41.7 & $+28.8$ & 30.1 & 72 / 1 / 27\%\\
Reference model, MSE loss & \textbf{71.4} & 67.6 & \textbf{5.1} & \textbf{22.4} & {\boldmath$+1.5$} & \textbf{22.3} & 90 / 1 / 10\%\\
Linear map & 79.4 & 67.6 & 6.9 & 41.0 & $+27.1$ & 30.8 & 73 / 1 / 27\%\\
NAM & 81.0 & 67.6 & 12.2 & 42.8 & $+26.8$ & 33.4 & 70 / 2 / 28\%\\
Four-coefficient model & 82.9 & 67.6 & 7.9 & 47.3 & $+34.8$ & 32.0 & 67 / 1 / 33\%\\
Four-curve model & 80.7 & 67.6 & 9.2 & 43.0 & $+28.4$ & 32.2 & 70 / 1 / 28\%\\\hline
\end{tabular}
\end{table}

Nearly three quarters of the reference model's squared error is noise given its inputs, and the variance from training is about 1\%. The
bias is the same for the network, the NAM and the linear map (41--43~km~s$^{-1}$ RMS), so the reducible error is common
to all model classes and is not a matter of flexibility. About half of the reference model's squared bias is the offset introduced by the
$S$ loss; it grows with the expected speed, from $+19$ to $+71$~km~s$^{-1}$ across the range of $\mathrm{E}[Y\mid X]$, as
expected from the outcome weighting (Section~\ref{sec:metric}). Without that offset, the MSE-trained reference model retains a shape
bias of 22~km~s$^{-1}$ (2--31~km~s$^{-1}$ across the interval of the noise estimate), so that 90\% of its squared error
is noise. This is the error that additional information or better physics of propagation and timing would have to
remove. The budget inherits the caveats of the noise estimate: it is relative to these inputs and to the reference model's
propagation, used for all models, and the variance is a lower bound because it omits the variation of the training set
across folds.

\section{Discussion and outlook}
\label{sec:discussion}
The results show that a map from solar observations to the inner-boundary speed can be learned from downstream measurements alone, by training it through a propagation to the spacecraft, and that the result is more skilful than both a conventional image-based regression and the WSA--HUXt chain. The factorial comparison identifies what makes this work. The essential element is that the map is trained through a propagation in which the boundary speed itself sets the travel time of each parcel, so that the predicted series is not a fixed time shift of the boundary; with a fixed lag, most of the gain is lost. The rest of the gain comes from the inputs and the training score. The capacity of the network adds nothing measurable: a linear function of the same inputs, trained in the same way, does as well. The differentiable formulation is what makes this training possible, without a hand-crafted matching between images and in-situ data.

The accuracy of the reconstruction is limited by the information in its inputs rather than by the model. About 57\% of the variance of the hourly speed is not determined by these inputs, and for about one fast stream in six, the most similar situations in other years were followed by slow wind. The part of the error that is not noise is nearly the same for the network, the additive models and the linear map (Table~\ref{tab:budget}), so a more flexible map would have little to gain. This agrees with the finding of \citeA{Collin2025} that simple models on coronal-hole areas match deep networks on full images. The noise is defined relative to the chosen features, so features learned directly from the full images could in principle reduce it. They would, however, raise the variance: the development interval contains only about 200 fast-wind streams, and even the flexible boundary maps tested here fit the training batches better without gaining on validation or test data. Substantial progress therefore requires inputs that carry additional information about the solar source and its connection to Earth.

The training and evaluation score is a second contribution of this paper. Training on the mean squared error compresses the predicted amplitude of rare fast streams, so we introduced TIMSE, the mean squared error over the hours faster than a threshold, integrated over all thresholds; it weights the error of each hour increasingly with its observed speed without fixing an event threshold. We trained on $S$, the geometric mean of TIMSE and the mean squared error. Compared with the same network trained on the mean squared error, $S$ buys amplitude in fast wind and detection at fixed thresholds at the price of a positive bias, while the ability to rank fast wind above slow wind is nearly unchanged. Because its weights depend on the observed outcome, $S$ rewards over-forecasting, a known property of outcome-weighted scores \cite{Gneiting2011,Lerch2017}; the choice of $S$ is therefore a statement about the application, and we report both models. Losses that target a specified statistic, such as quantile or expectile losses, or probabilistic training with threshold-weighted proper scores, would emphasize the fast tail without rewarding bias.

No single score captures both the bulk of the speed distribution and its fast tail. We therefore evaluated every model also by its errors conditioned on the observed and on the predicted speed, and by the tail-conditional RMSE as a function of the threshold (Figures~\ref{fig:condrmse} and~\ref{fig:condrmse_links}). These curves show where a change of training score or of propagation helps or hurts, which global scores hide, and they reveal effects, such as the levelling-off of speed with very large coronal holes, that a global score can miss entirely. We propose them as a standard complement to global scores in the evaluation of solar-wind reconstructions and forecasts.

Within the tolerance that ACE allows, the boundary map is close to linear in four physical quantities, dominated by coronal-hole area as in the empirical relations on which ESWF is built. A model with ten weights that comes within the uncertainty of the network is useful in its own right: its coefficients can be checked against physical expectation and carried to other periods and instruments.

What, then, is the reconstructed boundary? It is not a measurement of the solar-wind speed at $21.5\,R_\odot$, but an effective condition: the boundary that, propagated by this particular model, reproduces the speeds observed at ACE. Three things limit how far it can be read as a physical speed. First, ACE sees the boundary only along the sub-Earth track; everywhere else the boundary is the map's prediction from the same kind of inputs, not a fit to data. Second, even along the track the data leave some freedom: boundary maps that fit ACE equally well, such as two training runs of the same model or the network and the linear map, differ by 16--23~km~s$^{-1}$ RMS. Third, the absolute level of the boundary trades off against the residual acceleration of the propagation model, which ACE constrains only as a net speed added between the boundary and the spacecraft (Section~\ref{sec:autodiff}). Comparisons of the reconstructed boundary with WSA or other boundary speeds should therefore allow for an offset, and the boundary should be used with the propagation it was trained through (Section~\ref{sec:inverse}). The way to constrain it further is to add observers: a spacecraft at another longitude samples a second track, which the co-rotating formulation accommodates directly, and Part~II uses STEREO-A, Parker Solar Probe and Solar Orbiter to test the boundary away from the Sun--Earth line.

\section{Conclusions}
\label{sec:conclusions}
We have posed the reconstruction of the solar-wind speed at $21.5\,R_\odot$ as an inverse problem, in which a map from solar observations to the boundary is trained against ACE measurements through a time-dependent heliospheric transport model. The framework is conveniently expressed as an end-to-end differentiable optimization: the boundary is never observed, and the map learns it only through the physical model that links it to the measurements, which makes the approach a form of self-supervised, physics-informed machine learning. Five conclusions follow from the development data of solar cycle 24.
First, coronal-hole area dominates the information that 193~\AA{} images carry about the wind; its proximity to the ecliptic and its darkness add to it, the speed observed one rotation earlier adds most beyond the images, and HMI magnetic summaries add nothing measurable.
Second, trained through the propagation, the boundary map outperforms 27-day recurrence, a linear regression of the image features at a fixed lead and a GONG-driven WSA model propagated with HUXt, with a small but resolved gain in the discrimination of fast wind (AUC at 600~km~s$^{-1}$ 0.879 versus 0.850) and a much larger amplitude (0.81 versus 0.57), at the cost of a $+26$~km~s$^{-1}$ bias set by the tail-weighted loss.
Third, the gain comes from training through a propagation in which the travel time depends on the speed. Resolving the interaction of fast and slow streams, that is using the transport solver instead of ballistic propagation, adds little, and so does the capacity of the network: a linear map of the same inputs, trained in the same way, does as well as the network, and a model with one coefficient per physical quantity comes within its uncertainty.
Fourth, about 57\% of the variance of the hourly speed is not determined by these inputs, and the remaining error is common to all model classes, so that better reconstructions need more informative inputs rather than more flexible maps.
Fifth, the reconstructed boundary is an effective, model-dependent condition: ACE constrains it along one track, to within about 16--23~km~s$^{-1}$, and only up to an offset shared with the residual acceleration.
Forecasting at fixed issue times, with an independent test on cycle-25 data, and calibrated probabilistic prediction are the subjects of Parts~II and~III.

\section*{Open Research Section}
Upon submission.

\acknowledgments

\appendix
\section{Numerical propagation and differentiability}
\label{app:numerics}

The reduced transport equation is discretized on a uniform longitude--radius--time mesh, $r_i=r_0+i\Delta r$, $\phi_j=j\Delta\phi$, and $t^n=n\Delta t$, with periodic longitude and cell averages $v_{i,j}^n$. The inner-boundary condition is imposed as $v_{0,j}^n=u_j^n$. A steady-state initializer neglects the explicit time derivative and remaps each radial profile according to the travel-time shift $\Delta\phi_i=\Omega\Delta r/v_r$; values falling between longitude cells are obtained by periodic interpolation.

The time-dependent equation (Equation~\ref{eq:forward_accel}) is written in conservative form,
\begin{equation}
\frac{\partial v_r}{\partial t}+\frac{\partial}{\partial r}\left(\frac{v_r^2}{2}\right)-\frac{\partial}{\partial\phi}(\Omega v_r)=v_r\frac{\mathrm{d}\Delta V}{\mathrm{d}r}.
\label{eq:forward_conservative}
\end{equation}
Spatial fluxes use finite-volume MUSCL reconstruction \cite{vanLeer1979} with the monotonized-central limiter \cite{vanLeer1977}; see \citeA{LeVeque2002} for a general account. For adjacent cell differences $\delta^+v_{i,j}=v_{i+1,j}-v_{i,j}$ and $\delta^-v_{i,j}=v_{i,j}-v_{i-1,j}$, define
\begin{equation}
\rho_{i,j}=\frac{\delta^-v_{i,j}}{\delta^+v_{i,j}+\varepsilon},\qquad
\psi(\rho)=\max\left(0,\min\left(2\rho,\frac{1+\rho}{2},2\right)\right),\qquad
s_{i,j}=\psi(\rho_{i,j})\delta^+v_{i,j},
\end{equation}
where $\varepsilon>0$ prevents division by zero. Radial face states are $v^L_{i+1/2,j}=v_{i,j}+s_{i,j}/2$ and $v^R_{i+1/2,j}=v_{i+1,j}-s_{i+1,j}/2$. The radial Godunov flux uses $F_r(v)=v^2/2$; since the speed is always positive, it reduces to $F_r(v^L_{i+1/2,j})$, the reconstructed state of the upwind cell. Longitude uses the same limited reconstruction. The longitudinal flux is $F^\phi=-\Omega v$, so information travels toward decreasing Carrington longitude, and the upwind state at face $j+1/2$ is the reconstructed value from cell $j+1$: $F^\phi_{i,j+1/2}=-\Omega\,v^{R}_{i,j+1/2}$. The semi-discrete update is
\begin{equation}
\frac{\mathrm{d}v_{i,j}}{\mathrm{d}t}=L(v)_{i,j}=-\frac{F^r_{i+1/2,j}-F^r_{i-1/2,j}}{\Delta r}
-\frac{F^\phi_{i,j+1/2}-F^\phi_{i,j-1/2}}{\Delta\phi}
+v_{i,j}\frac{\mathrm{d}\Delta V}{\mathrm{d}r}\bigg|_{r_i}.
\label{eq:semi_discrete}
\end{equation}

Time integration is unsplit: the radial and longitude terms are advanced together by the two-stage strong-stability-preserving Runge--Kutta (Heun) method \cite{ShuOsher1988,Gottlieb2001}, $v^*=v^n+\Delta t\,L(v^n)$ and $v^{n+1}=v^n+(\Delta t/2)[L(v^n)+L(v^*)]$, and the boundary row $v_{0,j}=u_j$ is reimposed after each stage. As numerical safeguards the radial and longitude tendencies are each limited to $\pm50$~km~s$^{-2}$ and their sum to $\pm80$~km~s$^{-2}$, several orders of magnitude above physical tendencies, and the speed is limited to 100--900~km~s$^{-1}$ after each stage; these limits are part of the differentiated map, and gradients vanish where they are active.

The grid uses $\Delta r=(217-21.5)/63=3.10\,R_\odot$, $\Delta\phi=360^\circ/128$ and $\Delta t=1800$~s. The 128 longitude cells are periodic, so they span the full $360^\circ$ with spacing $360^\circ/128$; the 64 radial points include both ends of the domain, hence the divisor 63. An explicit scheme is stable only if the Courant number, the distance a signal travels in one time step in units of the grid spacing, is below one in each direction. With the Carrington sidereal rate $\Omega=2.8653\times10^{-6}$~rad~s$^{-1}$ (25.38~d), the longitude Courant number is $\Omega\Delta t/\Delta\phi=0.11$ (limit $\Delta t\leq17{,}131$~s), and the radial Courant number $v\Delta t/\Delta r$ is 0.50 at 600~km~s$^{-1}$ and 0.75 at 900~km~s$^{-1}$, the upper speed limit (limit $\Delta t\leq2399$~s). A stronger condition guarantees that the limited scheme creates no new extrema: for limiters bounded like the monotonized-central limiter, the forward-Euler step, and with it the two-stage strong-stability-preserving step, has this total-variation-diminishing property for Courant numbers up to 1/2 \cite{LeVeque2002,Gottlieb2001}. It holds in longitude and, in the radial direction, for speeds up to 600~km~s$^{-1}$; it is exceeded only in the fastest wind, where the verification below shows no spurious oscillations. Since the limiter is piecewise-defined, the discrete map is differentiable within each limiter branch, with nondifferentiable transitions at branch changes.

\begin{figure}[t]
\centering\includegraphics[width=\textwidth]{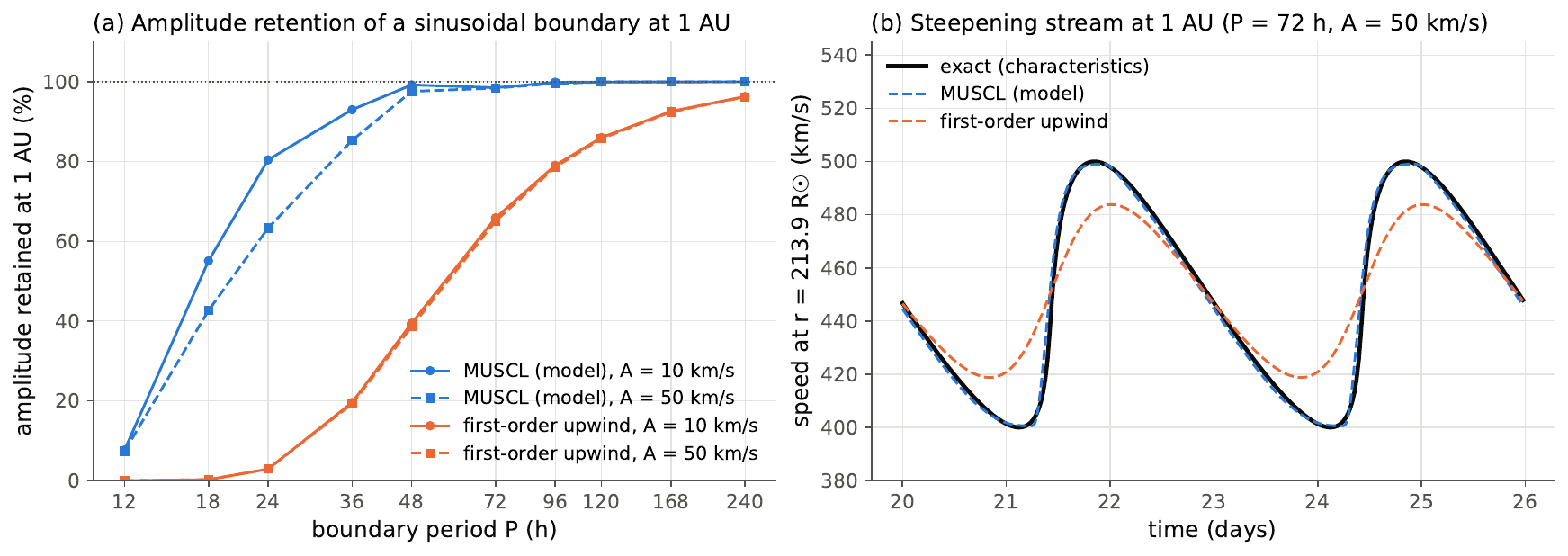}
\caption{Verification of the propagation solver on the model grid, residual acceleration off, for a boundary speed $450+A\sin(2\pi t/P)$~km~s$^{-1}$ uniform in longitude. (a) Fraction of the amplitude retained at 1~AU against the period, for the MUSCL scheme used in the model and a first-order upwind scheme, at $A=10$ and 50~km~s$^{-1}$. (b) Speed at 1~AU for $P=72$~h, $A=50$~km~s$^{-1}$: exact solution by characteristics, MUSCL and first-order upwind.}
\label{fig:verification}
\end{figure}

Figure~\ref{fig:verification} verifies the solver for a longitude-independent sinusoidal boundary, for which Equation~\ref{eq:forward_accel} without the acceleration term reduces to the inviscid Burgers equation. With the MUSCL fluxes, 100\%, 99\% and 80\% of a 10~km~s$^{-1}$ boundary amplitude reach 1~AU at periods of 120, 48 and 24~h, compared with 86\%, 39\% and 3\% for first-order upwind fluxes on the same grid; the 0.5--3-day band that carries stream structure is therefore transmitted. For a 72-h period and 50~km~s$^{-1}$ amplitude, the stream steepens without forming a shock before 1~AU, and the solver reproduces the exact solution obtained by characteristics with an RMS difference of 3.2~km~s$^{-1}$ (first-order: 14.6~km~s$^{-1}$).

\section{Data processing details}
\label{app:data}
During large flares, automatic exposure control shortens exposures below 0.1~s, and such frames produce spuriously large dark areas; 28 affected development-window hours were corrected. For nine hours, a normal-exposure frame within $\pm10$~min was substituted; for the other 19, AIA features and visibility metadata from the preceding hour were carried forward, without dropping hourly time steps.

GONG produced the synoptic maps every 6~h until October 2012 and about hourly afterwards; the median age of the map used is 3.1~h before October 2012 and 0.9~h afterwards. All 60,181 maps needed for the development window were extrapolated successfully, and every hour has a value. A median of 126 of the 128 field lines per map return a valid expansion factor (5th percentile 118). Over the development window, $\log_{10}f_s$ has a median of 2.0 and a 5--95\% range of 1.1--3.3.

\begin{figure*}[t]
\centering\includegraphics[width=\textwidth]{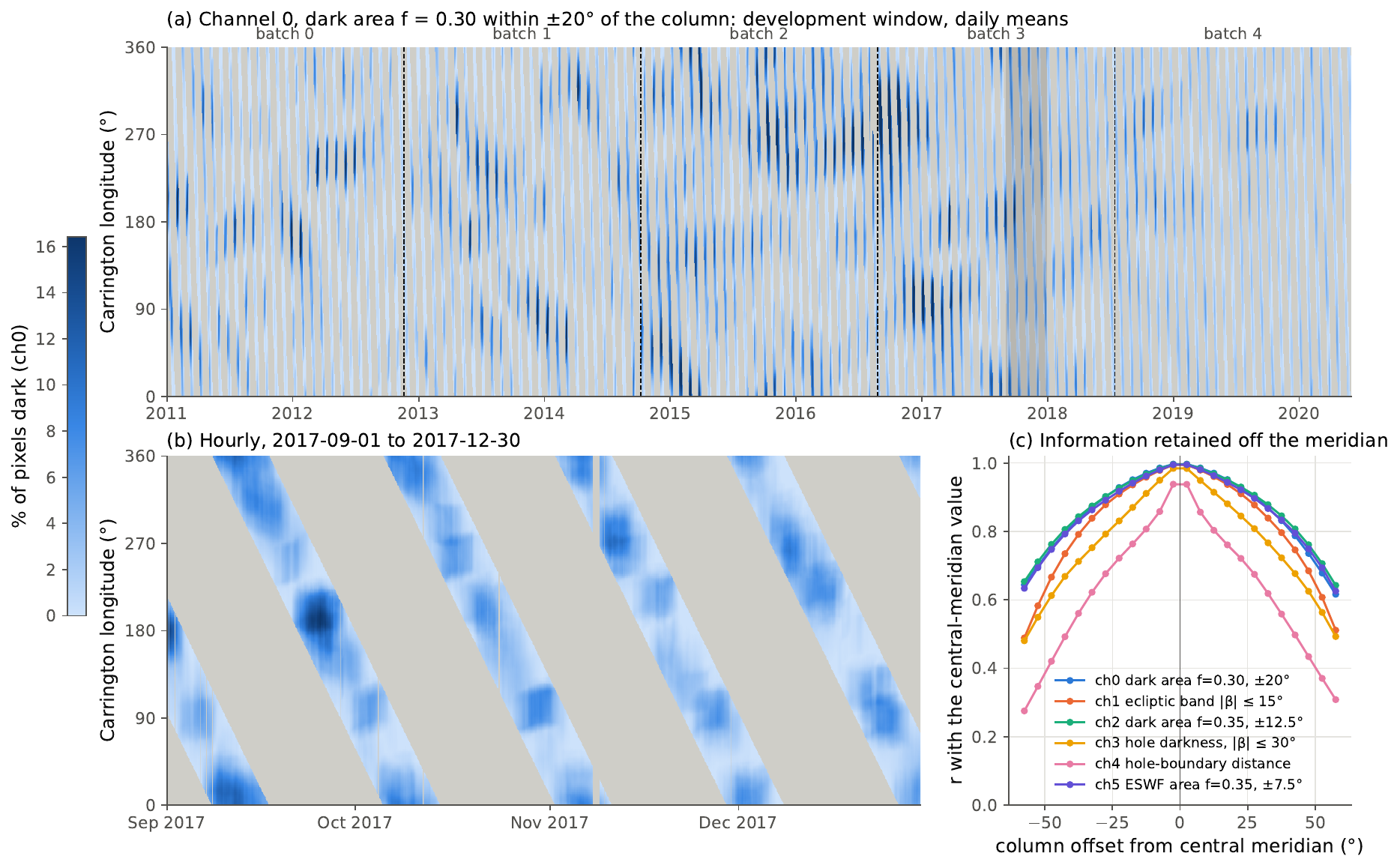}
\caption{Longitude-resolved solar constraints and view-dependent information loss. (a) Channel 0 (dark area, $f=0.30$, half-width $20^\circ$) as a time--Carrington-longitude map over the development window (daily means of visible values; dashed lines: batch boundaries); (b) the same channel hourly over 120 days, showing the visible longitude band drifting with solar rotation (grey: not visible or missing). (c) Correlation between each AIA channel measured at a given offset from the central meridian and the same Carrington longitude measured at its meridian crossing.}
\label{fig:geometry}
\end{figure*}

Figure~\ref{fig:geometry} shows how the AIA features are observed as the visible longitude band drifts with solar rotation, and how their information degrades away from the central meridian.

\section{Feature screening}
\label{app:screening}
This appendix gives the details of the feature screening of Section~\ref{sec:screening}.

\begin{figure*}[t]
\centering\includegraphics[height=0.68\textheight]{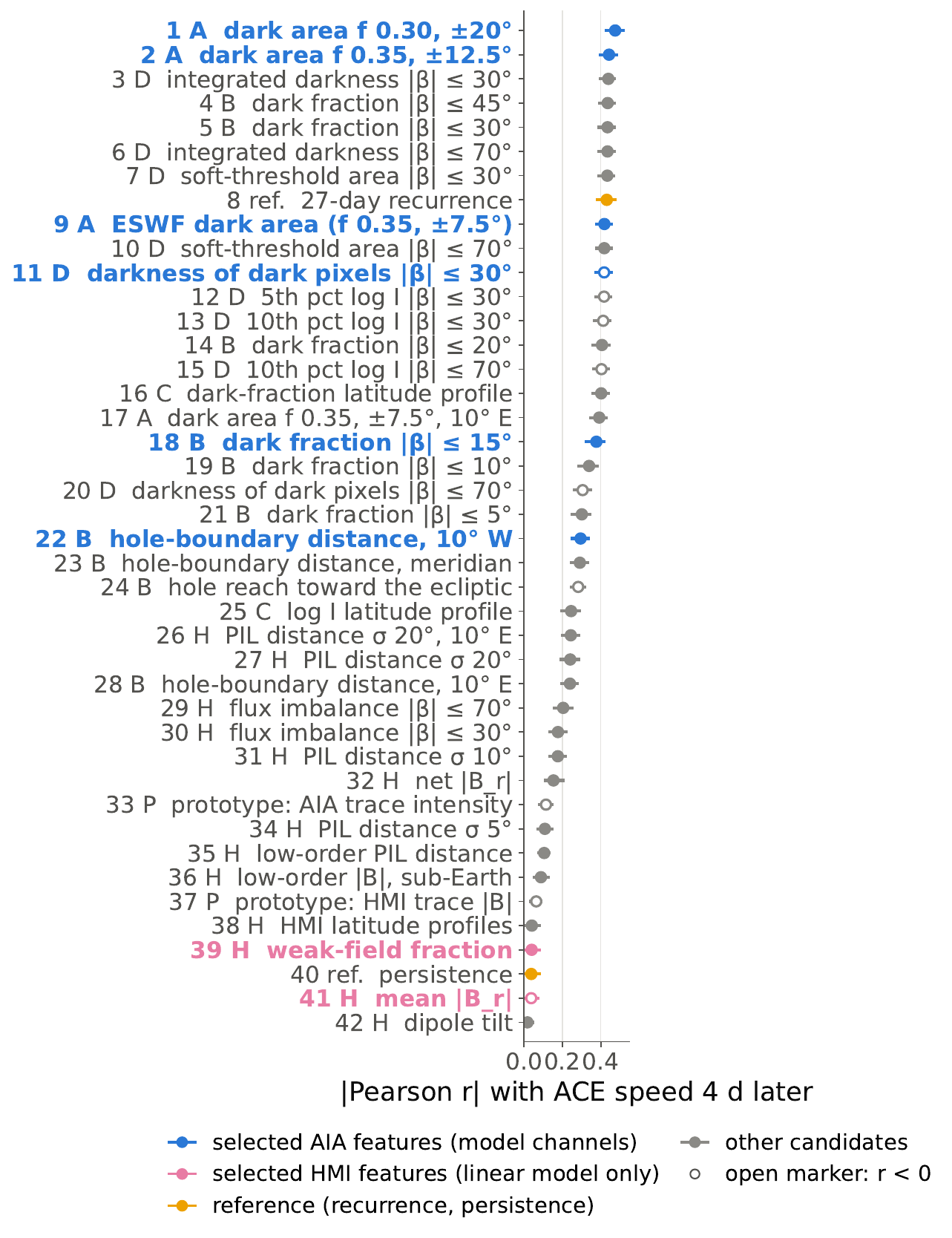}
\caption{Predictive information of the 42 candidate features (development data; target: ACE speed four days after the
image). Absolute Pearson correlation, mean of the five within-batch values, with the 95\% rotation-block bootstrap
interval; open markers denote negative correlations. Labels give the rank, the family and a short name; full definitions
are in Table~\ref{tab:allfeatures}, in the same order. The six AIA features of Table~\ref{tab:channels} are in blue, the
two HMI features used only by the linear comparator in pink, and the recurrence and persistence references in yellow.}
\label{fig:ranking}
\end{figure*}

The uncertainty of each correlation is estimated with a block bootstrap. The pairs are divided into consecutive blocks of one solar rotation (27.275~days); a resample is formed by drawing as many blocks as there are, at random and with replacement, and the score is recomputed on it. Repeating this 2,000 times gives the distribution of the score, and its 2.5th and 97.5th percentiles form the 95\% interval. We resample whole rotations rather than individual hours because consecutive hours are strongly correlated, and treating them as independent would give intervals that are too narrow. A few candidates are latitude profiles, vectors with one value per $5^\circ$ latitude bin, rather than single numbers. For these, a ridge regression is fitted to predict the ACE speed four days later from the profile, using the other four batches, and the correlation between its predictions and the observed speed on the held-out batch is the batch score; as for the other candidates, the five batch scores are averaged. Figure~\ref{fig:ranking} ranks all 42 candidates, and Table~\ref{tab:allfeatures} defines them in the same order.

\begin{figure*}[t]
\centering\includegraphics[width=\textwidth]{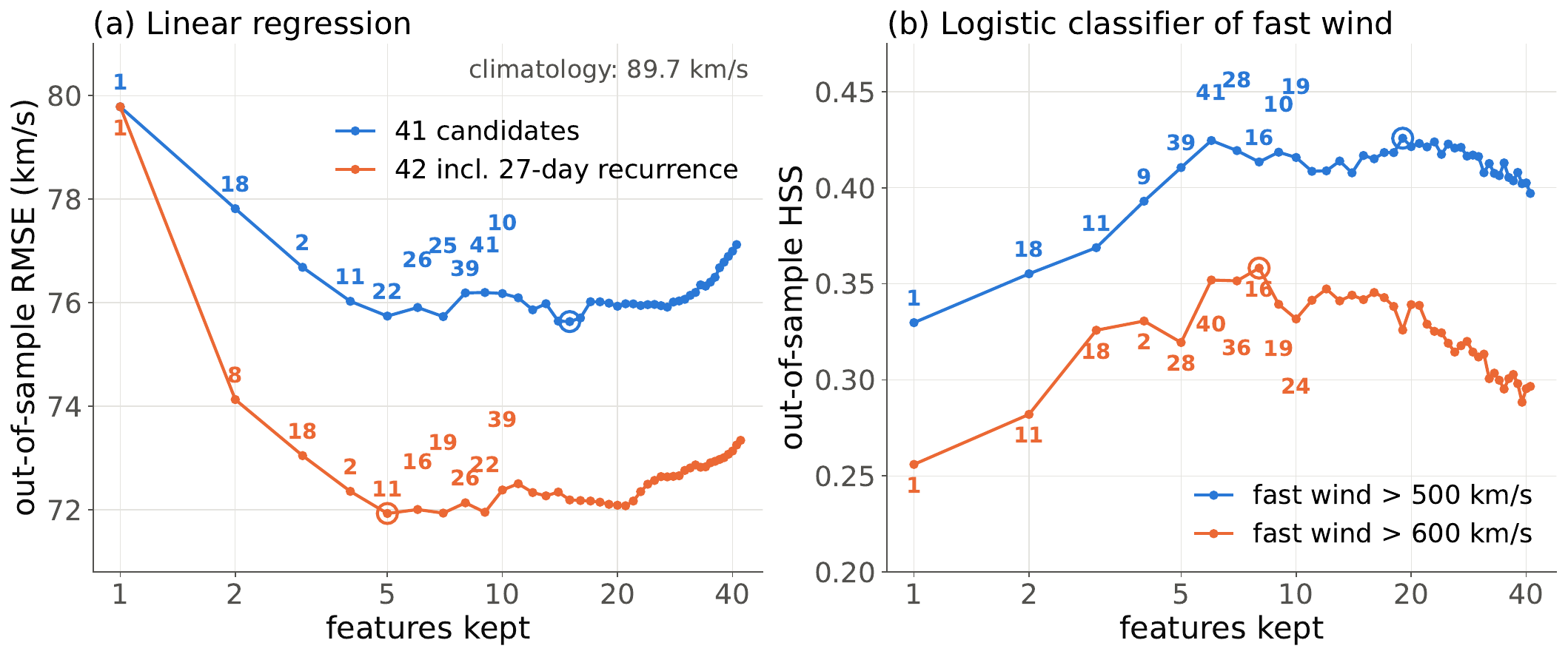}
\caption{Backward elimination on the development data: at each step the candidate with the smallest absolute $t$
(regression) or Wald (classifier) statistic is removed, and the model is scored on the held-out batches. (a)
Out-of-sample RMSE of a linear regression, for the candidates without and with the 27-day recurrence; circles mark the
minima. (b) Heidke skill score of a logistic classifier of fast wind ($v>500$ and $v>600$~km~s$^{-1}$); circles mark
the maxima. The numbers at the first ten points of each path are the ranks (Figure~\ref{fig:ranking}) of the feature that
enters the model at that number of features.}
\label{fig:elimination}
\end{figure*}

In the backward elimination, the candidate removed at each step is the one with the smallest absolute $t$ statistic: the fitted regression coefficient divided by its standard error, which ordinary least squares provides directly from the scatter of the residuals and from how much the feature varies independently of the other features. The feature with the smallest $|t|$ adds least, given the others. These statistics only order the removals and are not used as significance tests. Along the whole path, the out-of-sample RMSE stays well below the 89.7~km~s$^{-1}$ of climatology (Figure~\ref{fig:elimination}a). The classifier version predicts, from the same standardized candidates, the probability that the speed four days later exceeds 500 or 600~km~s$^{-1}$ (16\% and 5\% of the hours), and removes at each step the candidate with the smallest absolute Wald statistic, the logistic counterpart of the $t$ statistic. Each classifier is scored on the held-out batches by the Heidke skill score (HSS), the fraction of correct fast/not-fast forecasts in excess of that expected by chance, with fast wind forecast when the predicted probability exceeds the threshold that maximizes the HSS on the training batches. At 500~km~s$^{-1}$ the HSS is nearly flat from six features on (0.40--0.43); at 600~km~s$^{-1}$ it peaks with eight features (0.36) and declines as more are added, because so few hours exceed that speed (Figure~\ref{fig:elimination}b).

\footnotesize
\begin{longtable}{rlp{0.58\linewidth}cc}
\caption{All candidate features screened (Section~\ref{sec:screening}), in the order of Figure~\ref{fig:ranking}. Families: A, dark area in a full-height strip; B, proximity of dark regions to the ecliptic; C, latitude profiles; D, intensity within the dark regions; H, HMI magnetic field; ref., reference predictors. Unless stated otherwise, ``dark'' means $I<0.35\,I_{\mathrm{med}}$, the central strip is $|\Delta\mathrm{lon}|\leq7.5^\circ$, and full-height strips span $|\beta|\leq70^\circ$; coronal holes are connected dark regions of at least 5~deg$^2$. $r$: Pearson correlation with the ACE speed four days after the image, mean of the five within-batch values, with the 95\% rotation-block bootstrap interval. Use: channel of the reference model (Table~\ref{tab:channels}), or Lin.\ for the two HMI features used only by the linear comparator.}\label{tab:allfeatures}\\
\hline
 & Fam. & Definition & $r$ [95\% interval] & Use\\\hline
\endfirsthead
\caption[]{All candidate features screened (continued)}\\
\hline
 & Fam. & Definition & $r$ [95\% interval] & Use\\\hline
\endhead
\hline
\endfoot
\hline
\endlastfoot
1 & A & Dark area ($f=0.30$), strip of half-width $20^\circ$ centred on the meridian & $+0.47$ [$+0.43$, $+0.52$] & Ch.~0\\
2 & A & Dark area ($f=0.35$), strip of half-width $12.5^\circ$ centred on the meridian & $+0.44$ [$+0.40$, $+0.48$] & Ch.~2\\
3 & D & Integrated darkness $\sum\max(0,\log_{10}0.35-\log_{10}I/I_{\mathrm{med}})/N$, central strip, $|\beta|\leq30^\circ$ & $+0.44$ [$+0.40$, $+0.47$] & \\
4 & B & Dark fraction within $|\beta|\leq45^\circ$, central strip & $+0.44$ [$+0.39$, $+0.47$] & \\
5 & B & Dark fraction within $|\beta|\leq30^\circ$, central strip & $+0.43$ [$+0.39$, $+0.47$] & \\
6 & D & Integrated darkness $\sum\max(0,\log_{10}0.35-\log_{10}I/I_{\mathrm{med}})/N$, central strip, $|\beta|\leq70^\circ$ & $+0.43$ [$+0.39$, $+0.47$] & \\
7 & D & Dark area with a logistic (soft) threshold at $0.35\,I_{\mathrm{med}}$, central strip, $|\beta|\leq30^\circ$ & $+0.43$ [$+0.39$, $+0.47$] & \\
8 & ref. & Reference: ACE speed one solar rotation (27~d) before the target hour & $+0.43$ [$+0.38$, $+0.47$] & \\
9 & A & Dark area ($f=0.35$), half-width $7.5^\circ$ (ESWF coronal-hole area) & $+0.42$ [$+0.38$, $+0.45$] & Ch.~5\\
10 & D & Dark area with a logistic (soft) threshold at $0.35\,I_{\mathrm{med}}$, central strip, $|\beta|\leq70^\circ$ & $+0.42$ [$+0.38$, $+0.45$] & \\
11 & D & Mean $\log_{10}(I/I_{\mathrm{med}})$ of the dark pixels, central strip, $|\beta|\leq30^\circ$ & $-0.42$ [$-0.45$, $-0.38$] & Ch.~3\\
12 & D & 5th percentile of $\log_{10}(I/I_{\mathrm{med}})$, central strip, $|\beta|\leq30^\circ$ & $-0.42$ [$-0.45$, $-0.37$] & \\
13 & D & 10th percentile of $\log_{10}(I/I_{\mathrm{med}})$, central strip, $|\beta|\leq30^\circ$ & $-0.41$ [$-0.45$, $-0.37$] & \\
14 & B & Dark fraction within $|\beta|\leq20^\circ$, central strip & $+0.41$ [$+0.36$, $+0.44$] & \\
15 & D & 10th percentile of $\log_{10}(I/I_{\mathrm{med}})$, central strip, $|\beta|\leq70^\circ$ & $-0.40$ [$-0.44$, $-0.36$] & \\
16 & C & Dark fraction in $5^\circ$ latitude bins of the central strip (profile, ridge) & $+0.40$ [$+0.36$, $+0.44$] & \\
17 & A & Dark area ($f=0.35$), strip of half-width $7.5^\circ$ centred 10$^\circ$ east of the meridian & $+0.39$ [$+0.35$, $+0.43$] & \\
18 & B & Dark fraction within $|\beta|\leq15^\circ$, central strip & $+0.38$ [$+0.33$, $+0.42$] & Ch.~1\\
19 & B & Dark fraction within $|\beta|\leq10^\circ$, central strip & $+0.34$ [$+0.29$, $+0.38$] & \\
20 & D & Mean $\log_{10}(I/I_{\mathrm{med}})$ of the dark pixels, central strip, $|\beta|\leq70^\circ$ & $-0.31$ [$-0.35$, $-0.26$] & \\
21 & B & Dark fraction within $|\beta|\leq5^\circ$, central strip & $+0.30$ [$+0.25$, $+0.34$] & \\
22 & B & Signed distance from the ecliptic to the nearest hole boundary, 10$^\circ$ west of the meridian (positive inside a hole) & $+0.29$ [$+0.25$, $+0.34$] & Ch.~4\\
23 & B & Signed distance from the ecliptic to the nearest hole boundary, at the meridian (positive inside a hole) & $+0.29$ [$+0.25$, $+0.33$] & \\
24 & B & Smallest $|\beta|$ reached by a coronal hole in the central strip & $-0.28$ [$-0.32$, $-0.25$] & \\
25 & C & Mean $\log_{10}(I/I_{\mathrm{med}})$ in $5^\circ$ latitude bins of the central strip (profile, ridge) & $+0.25$ [$+0.20$, $+0.29$] & \\
26 & H & Distance from the ecliptic to the polarity inversion line of $B_r$ smoothed at $20^\circ$, 10$^\circ$ east of the meridian & $+0.24$ [$+0.20$, $+0.29$] & \\
27 & H & Distance from the ecliptic to the polarity inversion line of $B_r$ smoothed at $20^\circ$ & $+0.24$ [$+0.19$, $+0.29$] & \\
28 & B & Signed distance from the ecliptic to the nearest hole boundary, 10$^\circ$ east of the meridian (positive inside a hole) & $+0.24$ [$+0.20$, $+0.28$] & \\
29 & H & Flux imbalance $|\sum B_r|/\sum|B_r|$, half-width $20^\circ$, $|\beta|\leq70^\circ$ & $+0.20$ [$+0.16$, $+0.25$] & \\
30 & H & Flux imbalance $|\sum B_r|/\sum|B_r|$, half-width $20^\circ$, $|\beta|\leq30^\circ$ & $+0.18$ [$+0.13$, $+0.22$] & \\
31 & H & Distance from the ecliptic to the polarity inversion line of $B_r$ smoothed at $10^\circ$ & $+0.18$ [$+0.14$, $+0.22$] & \\
32 & H & Net flux $|\sum B_r|$, half-width $20^\circ$, $|\beta|\leq70^\circ$ & $+0.15$ [$+0.11$, $+0.21$] & \\
33 & ref. & Mean AIA intensity, $|\beta|\leq3^\circ$, central strip (input of an earlier prototype) & $-0.11$ [$-0.15$, $-0.08$] & \\
34 & H & Distance from the ecliptic to the polarity inversion line of $B_r$ smoothed at $5^\circ$ & $+0.11$ [$+0.07$, $+0.15$] & \\
35 & H & Inversion-line distance of the $l\leq2$ harmonic fit to the visible field & $+0.11$ [$+0.08$, $+0.13$] & \\
36 & H & $|B|$ of the $l\leq2$ harmonic fit at the sub-Earth point & $+0.09$ [$+0.05$, $+0.13$] & \\
37 & ref. & Mean HMI $|B_{\mathrm{los}}|$, $|\beta|\leq3^\circ$, central strip (input of an earlier prototype) & $-0.06$ [$-0.09$, $-0.03$] & \\
38 & H & $|B_r|$, net flux and imbalance in latitude bins (profiles, ridge) & $+0.04$ [$0.00$, $+0.08$] & \\
39 & H & Fraction of pixels with $|B_r|<10$~G, half-width $7.5^\circ$, $|\beta|\leq30^\circ$ & $+0.04$ [$0.00$, $+0.08$] & Lin.\\
40 & ref. & Reference: ACE speed at the image time (persistence) & $+0.04$ [$-0.01$, $+0.08$] & \\
41 & H & Mean $|B_r|$, half-width $7.5^\circ$, $|\beta|\leq30^\circ$ & $-0.04$ [$-0.07$, $0.00$] & Lin.\\
42 & H & Dipole tilt of the $l\leq2$ harmonic fit & $+0.02$ [$-0.01$, $+0.05$] & \\
\hline
\end{longtable}
\normalsize

\section{The training score}
\label{app:score}
\begin{figure}[th]
\centering\includegraphics[width=0.75\textwidth]{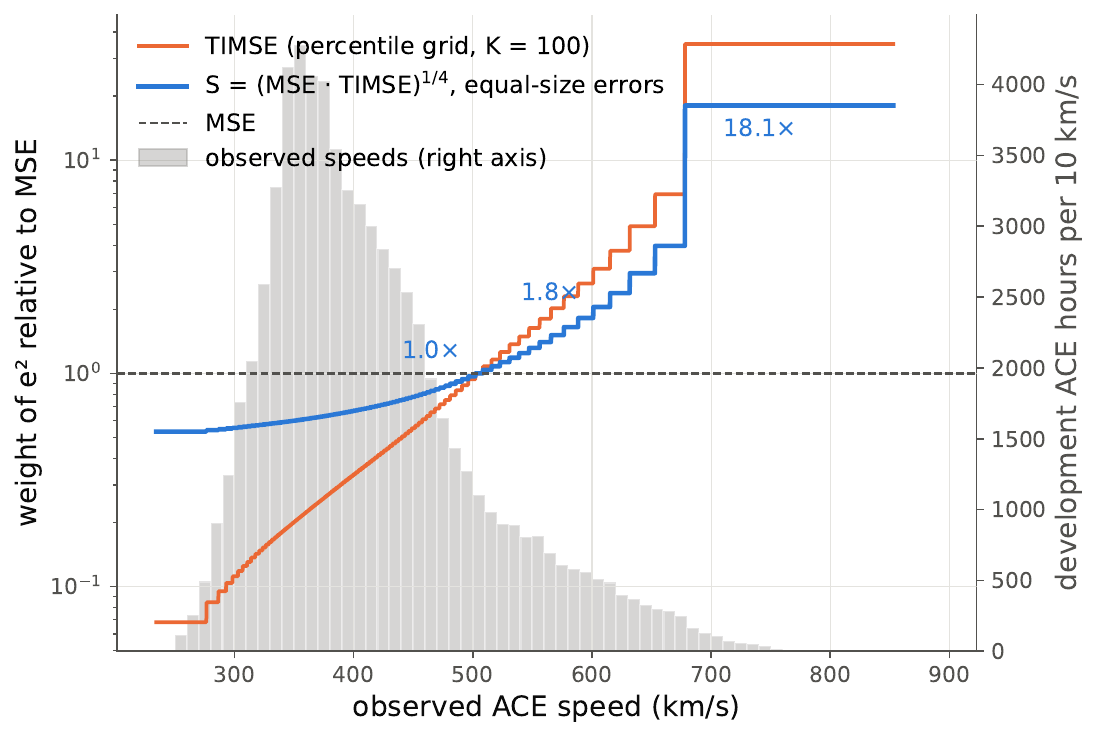}
\caption{What the score $S$ weights. Effective weight of the squared error of an hour with observed speed $y$, relative to MSE (dashed): $\operatorname{TIMSE}_N$ on the 100-percentile grid (orange) and $S$ for errors of equal size (blue). The steps are the percentile intervals; the last interval holds the top 1\% of hours. Grey, right axis: distribution of
the 70,351 development ACE hours (1 January 2011 to 1 June 2020) that define the grid, in 10~km~s$^{-1}$ bins.}
\label{fig:weights}
\end{figure}

This appendix shows how the score $S$ of Section~\ref{sec:metric} weights the error of each hour. Because $\operatorname{TIMSE}_N$ is a weighted sum of squared errors (Equation~\ref{eq:timse}), the weight of an hour's squared error depends only on its observed speed. Figure~\ref{fig:weights} shows this weight relative to the MSE, which weights all hours equally (dashed): normalized to an average of one, it ranges from below 0.1 for the slowest hours to about 35 in the fastest percentile (orange). The score $S$ combines the MSE and $\operatorname{TIMSE}_N$, and a small change in the squared error of one hour changes $S$ by the average of its effects on the two. For errors of equal size, the weights of $S$ are therefore the average of those of the MSE and of $\operatorname{TIMSE}_N$ (blue): 0.67 at 400~km~s$^{-1}$, one near 500~km~s$^{-1}$, 1.8 at 600~km~s$^{-1}$ and 18 in the top percentile of observed speeds (above about 680~km~s$^{-1}$). $S$ thus keeps every hour in the score while giving fast wind much more weight. The steps of the curves follow the percentile grid, whose spacing is set by the distribution of observed speeds (grey).

\newpage
\bibliography{references}
\end{document}